\documentclass[10pt,amsmath,amssymb,nofootinbib,twoside,twocolumn,superscriptaddress,floats,floatfix,aps,preprintnumbers,prd]{revtex4-2}
\usepackage[british]{babel}
\usepackage{verbatim}
\usepackage{latexsym}
\usepackage{graphicx}
\usepackage{tabularx}
\usepackage{cancel}
\usepackage{hyperref}
\usepackage{colortbl}
\usepackage{aas_macros}
\usepackage{multirow}
\usepackage{makecell}
\usepackage{booktabs}
\usepackage{hhline}
\usepackage[capitalise]{cleveref}
\usepackage[utf8]{inputenc}
\usepackage[normalem]{ulem}
\usepackage[dvipsnames]{xcolor}
\usepackage{subfigure}
\graphicspath{{corner_plots/}}
\usepackage{bm}

\usepackage{placeins}  
\newcommand{\dd}{{\rm d}}          
\newcommand{\tc}{\tilde{c}}      
\newcommand{\m}{{\rm m}}
\newcommand{\dS}{{\rm dS}}
\newcommand{\ec}{{\rm ec}}

\newcommand{\nn}{\nonumber}
\newcommand{\vf}{\varphi}

\begin{document}

\newcommand{\Salamanca}{\affiliation{Departamento de F\'isica Fundamental, Universidad de Salamanca, Plaza de la Merced, s/n, E-37008 Salamanca, Spain.}}
\newcommand{\SalamancaIUFFyM}{\affiliation{Instituto Universitario de F\'isica Fundamental y Matem\'aticas (IUFFyM), Universidad de Salamanca, Plaza de la Merced, s/n, E-37008 Salamanca, Spain.}}
\newcommand{\SSM}{\affiliation{{Scuola Superiore Meridionale, Via Mezzocannone 4,  I-80134 Napoli, Italy.}}}
\newcommand{\infn}{\affiliation{Istituto Nazionale di Fisica Nucleare (INFN), sezione di Napoli, via Cinthia 9, I-80126 Napoli, Italy.}}
\newcommand{\Napoli}{\affiliation{Dipartimento di Fisica ``E. Pancini", Università degli Studi di Napoli ``Federico II",  via Cinthia 9, I-80126 Napoli, Italy.}}
\newcommand{\Salerno}{\affiliation{Dipartimento di Fisica ``E.R. Caianiello'', Universit\`a di Salerno, Via Giovanni Paolo II 132,
I-84084 Fisciano, Italy.}}

\author{M.~Miranda}
\email{m.miranda@ssmeridionale.it}
\SSM
\infn

\author{Ruchika}
\email{ruchika.science@usal.es}
\Salamanca
\SalamancaIUFFyM

\author{I.~De~Martino}
\email{ivan.demartino@usal.es}
\Salamanca
\SalamancaIUFFyM

\author{D.~Vernieri}
\email{daniele.vernieri@unina.it}
\Napoli
\SSM
\infn
\Salerno

\author{S.~Capozziello}
\email{capozziello@na.infn.it}
\Napoli
\SSM
\infn

\title{Cosmological constraints and standard sirens forecasts for non-dynamical dark energy in Horndeski gravity}

\begin{abstract}
We investigate an analytically tractable sector of the Extended Cuscuton model, a non-dynamical dark-energy realization within the framework of viable Horndeski gravity. We focus on four benchmark submodels and constrain them with current background probes, namely cosmic chronometers, Type-Ia supernovae, and BAO, while imposing theoretical viability, Lunar Laser Ranging, and Big Bang Nucleosynthesis bounds. We then forecast third-generation bright-standard-siren constraints with Einstein Telescope and Cosmic Explorer networks, considering prompt-emission, afterglow, and kilonova counterparts. Current data already restrict the viable parameter space to small departures from $\Lambda$CDM and do not remove the calibration-driven offset between the CC+SN and CC+BAO determinations of $H_0$. In principle, future bright sirens substantially sharpen the constraints, especially for kilonova catalogues and extended detector networks. Across the forecast configurations, the relative uncertainty on $H_0$ remains below $13.18\%$ and can reach $0.21\%$ in the most constraining cases, while $\Omega_\Lambda$ is recovered at the percent level in the best cases. These results show that third-generation standard sirens can provide a precise complementary test of non-dynamical dark energy beyond $\Lambda$CDM.
\end{abstract}

\preprint{ET-0537A-26}

\maketitle


\section{Introduction}\label{sec:intro}

Gravitational-wave (GW) astronomy opened a new observational window on cosmology. Once a compact-binary event is associated with an electromagnetic counterpart, it can be used as a standard siren, giving direct access to the luminosity distance without relying on the usual cosmic distance ladder~\cite{Taylor:2011fs, Belgacem:2017ihm}. Standard sirens are therefore relevant not only for measuring the Hubble constant, $H_0$, and other cosmological parameters, but also because GWs probe gravity over cosmological distances and thus bear directly on the physics of cosmic acceleration~\cite{Belgacem:2018lbp}.

This question is particularly relevant in view of the persistent Hubble tension~\cite{2025PDU....4901965D}. If the mismatch between early- and late-time determinations of $H_0$ is not entirely due to unresolved systematics, then the minimal $\Lambda$ Cold Dark Matter ($\Lambda$CDM) description may be incomplete. This could point to a modification of the late-time expansion history, of the dark sector, or of the gravitational sector itself. From this perspective, standard sirens are valuable not only because they provide an independent route to $H_0$, but because they allow one to ask a more basic question: does the inferred cosmology remain robust once one moves beyond $\Lambda$CDM? For recent discussions, see, e.g., Refs.~\cite{Capozziello:2025lor, Capozziello:2025qmh, Chaudhary:2025bfs}. This fact is particularly relevant because current GW measurements are still far from the percent-level precision reached by the main early- and late-time electromagnetic probes, and therefore are not yet decisive in the Hubble-tension debate.

This is one of the reasons third-generation gravitational-wave detectors matter for observational cosmology. With current detectors, standard-siren cosmology is still limited by small-number statistics and by the difficulty of identifying enough events with reliable redshifts. The situation is expected to change with the Einstein Telescope (ET) in Europe~\cite{Punturo:2010zz} and Cosmic Explorer (CE) in the United States~\cite{Evans:2021gyd}. Their improved sensitivity, broader frequency coverage, and operation as a network should extend the accessible volume by orders of magnitude, push bright-siren observations to much higher redshift, and improve both localization and distance reconstruction~\cite{Maggiore:2019uih, Borhanian:2020ypi, 2026JCAP...03..081A}. It is important to distinguish between bright sirens, for which an electromagnetic counterpart identifies the source and provides a redshift, and dark sirens, where the redshift is obtained only statistically. In the present work, the bright-siren channel is the relevant one, because the GW waveform alone measures redshifted masses and therefore does not provide an independent source redshift. An electromagnetic counterpart is what makes a direct cosmological use of the event possible. In practical terms, the step from current detectors to ET and CE is not only quantitative but it also changes what can realistically be extracted from the data. The point is not only that the error bars on $H_0$ become smaller. Once one moves beyond $\Lambda$CDM, parameter degeneracies broaden, and the constraining power of distance data degrades. The role of third-generation (3G) networks is therefore to make standard-siren cosmology informative precisely in the regime where the underlying model is more complex than a pure cosmological constant.

Standard sirens are also not just a probe for the background expansion. In many scalar-tensor theories, even when GWs propagate at the speed of light, their amplitude is modified by an additional friction term during propagation. As a result, the GW luminosity distance does not, in general, coincide with the electromagnetic one~\cite{Belgacem:2017ihm, Belgacem:2018lbp}. This means that standard sirens are sensitive both to the background history and to the tensor sector of the theory. This point is especially important for the present work, because in the considered theory of gravity, the same non-minimal coupling that affects the late-time cosmological dynamics also enters the tensor sector through the effective Planck mass. For dark-energy models beyond $\Lambda$CDM, this is precisely the kind of information that is needed~\cite{Lagos:2019kds}.

Among modified-gravity theories, Horndeski gravity has long provided the natural starting point~\cite{Horndeski:1974wa, Deffayet:2011gz, Kobayashi:2011nu}. It is the most general scalar-tensor theory with manifestly second-order equations of motion and has provided the standard framework for discussing dark energy, screening, cosmological perturbations, and GW phenomenology within a unified language~\cite{Kobayashi:2019hrl}. Before GW170817, the corresponding space of viable late-time models was broad. The joint observation of GW170817 and GRB170817A changed that picture sharply~\cite{LIGOScientific:2017vwq, LIGOScientific:2017ync}. Under the standard assumptions used in late-time dark-energy applications, the near equality between the speed of GWs and the speed of light at low redshift excludes large portions of Horndeski and beyond-Horndeski parameter space, leaving a much narrower class of surviving models~\cite{Baker:2017hug, Creminelli:2017sry, Ezquiaga:2017ekz, Langlois:2017dyl, Sakstein:2017xjx}. Models can also be selected on first-principle grounds, for instance by imposing symmetry requirements~\cite{Capozziello:2018gms, Miranda:2024hhe}. However, comparison with data remains essential for identifying observationally viable cosmologies.

Even within this reduced landscape, however, viability is not automatic. A model that aims to describe late-time acceleration must do more than fit a low-redshift Hubble diagram. It must be free from obvious instabilities, maintain a positive effective gravitational coupling, and remain compatible with structure formation and precision cosmology. For non-minimally coupled models, local tests are especially restrictive because the cosmological evolution of the effective gravitational strength can leave measurable traces even when screening mechanisms are active. Lunar Laser Ranging therefore provides an important bound on the present-day variation of Newton's constant~\cite{Williams:2004qba, Hofmann:2018myc}. Early-Universe physics provides a similar constraint: Big Bang Nucleosynthesis limits departures of the effective gravitational coupling from its present value through the expansion rate during primordial element formation~\cite{Alvey:2019ctk, Pitrou:2018cgg, Fields:2019pfx}. Any realistic dark-energy model must satisfy this full set of requirements, not only the late-time background tests.

Within this restricted class of theories, Extended Cuscuton is of particular interest because it reduces the scalar-sector dynamics without necessarily collapsing back to General Relativity. The scalar field does not introduce an additional propagating scalar mode, so that only the two tensor polarizations are dynamical~\cite{Iyonaga:2018vnu, Iyonaga:2020bmm}. Even so, the theory still changes cosmological evolution through a non-minimal coupling, a kinetic gravity braiding, and a non-trivial background scalar configuration, and can therefore play the role of non-dynamical dark energy. It thus provides a minimal and well-defined realization of non-dynamical dark energy within viable Horndeski gravity.

In Extended Cuscuton, the same sector that modifies the cosmological evolution also enters the effective Planck mass relevant for tensor propagation. Standard sirens are therefore sensitive not only to $H(z)$, but also to the running of the gravitational coupling felt by tensor modes~\cite{Belgacem:2018lbp, Lagos:2019kds}. This is precisely what makes the model phenomenologically interesting: any departure from $\Lambda$CDM is tied to a tightly constrained modification of gravity rather than to a broad extra-field phenomenology.

For this reason, in the present paper, we focus on a restricted set of Extended Cuscuton submodels selected according to \textit{simplicity criteria}. More precisely, we consider a subclass in which the scalar field can be solved algebraically from its equation of motion, which acts as a constraint, in close analogy with the construction discussed in~\cite{Iyonaga:2020bmm}. Then we focus on specific benchmark submodels having a non-degenerate $\Lambda$CDM limit. 

This non-dynamical sector also admits a formal Eckart-like effective-fluid interpretation~\cite{Giusti:2021sku, Miranda:2022wkz, Miranda:2024dhw, Gallerani:2024gdy}. Moreover, such Extended Cuscuton models can be used as an effective source for inhomogeneities embedded in a dynamical cosmological background~\cite{Afshordi:2014qaa, Miranda:2022brj} or to remove the classical Big Bang initial singularity and replace it with a non-singular bounce occurring at a critical energy density value~\cite{Miranda:2024aum}. These broader connections are not required for the analysis itself, but they help place the subclass studied here within a wider non-dynamical sector with a consistent physical interpretation.

The analysis developed in this paper is structured in two parts. First, in Sec.~\ref{sec:realdata}, we use current background probes, including CC, BAO, and SNeIa data, to determine which regions of parameter space remain viable once the background evolution model satisfies consistency requirements, {\em i.e.} positivity of the effective gravitational coupling, Lunar Laser Ranging bounds, and Big Bang Nucleosynthesis constraints. Second, in Sec.~\ref{sec:mockdata}, we use mock standard-siren catalogues to explore how future third-generation GW detector networks, in particular configurations involving ET and CE, can improve the sensitivity to these models through both the expansion history and the modified propagation of GWs. On the forecast side, we follow the multimessenger strategy that has become standard in recent 3G dark-energy studies, based on bright binary-neutron-star sirens with gamma-ray, X-ray, or kilonova counterparts and on explicit comparisons between ET, ET+CE, and ET+2CE network configurations~\cite{Califano:2022cmo, Califano:2023fbq, Califano:2024xzt}. The goal is to test whether a viable non-dynamical dark-energy model with only two propagating tensor modes, but with a modified effective Planck mass and hence a modified GW luminosity distance, remains observationally distinguishable from $\Lambda$CDM once bright-siren data from next-generation detectors are considered.

The layout of the paper is the following. In Sec.~\ref{sec:ecm}, analytical Extended Cuscuton models are discussed. In particular, we analyse the viability conditions and models with non-degenerate $\Lambda$CDM limit. In Sec.~\ref{sec:realdata}, cosmological data are taken into account. Specifically, we discuss the cosmological dataset, parameter priors, and the $\chi^2$ minimisation. Constraints coming from CC, BAO, and SNeIa are reported in Sec.~\ref{constraints}. The GW mock data and the data analysis procedure are reported in Sec.~\ref{sec:mockdata}. The cosmological parameter estimates as well as the fitting fiducial model are discussed in  Sec.~\ref{sec:results}. Possible forecasts with 3G Gravitational Waves detectors are obtained in Sec.~\ref{forecast}. Conclusions are drawn in Sec.~\ref{conclusions}.

\section{Analytical Extended Cuscuton}\label{sec:ecm}

Extended Cuscuton is a special subclass of scalar-tensor theories in which the scalar field contributes to the cosmological dynamics without introducing an independent propagating scalar mode~\cite{Iyonaga:2018vnu, Iyonaga:2020bmm}. The key point is not that the scalar disappears from the dynamics, but that its equation of motion becomes non-dynamical. On cosmological backgrounds, the structure of the action allows one to eliminate simultaneously the terms proportional to $\dot H$ and $\ddot\phi$ from the scalar-field equation, so that the latter acts as a constraint rather than as an independent second-order evolution equation. The gravitational sector therefore propagates only the two tensor polarizations of the metric, with the degree-of-freedom counting discussed in Ref.~\cite{Iyonaga:2020bmm}. In the presence of matter, the scalar mode that survives in cosmological perturbations is the usual matter one, while the Extended Cuscuton sector itself remains non-dynamical. Extended Cuscuton therefore provides a minimal realization of non-dynamical dark energy.

The model is also naturally compatible with the viable post-GW170817 Horndeski sector~\cite{Creminelli:2017sry, Ezquiaga:2017ekz, Sakstein:2017xjx, Langlois:2017dyl}. In Horndeski gravity, imposing that tensor modes propagate at the speed of light, $c_T=1$, on cosmological backgrounds removes the quartic and quintic structures that would modify the tensor speed, leaving a surviving action of the schematic form $\mathcal{L}=G_4(\phi)R+G_2(\phi,X)-G_3(\phi,X)\Box\phi$. The subclass considered here lies within this viable Horndeski sector, so tensor modes propagate luminally. This is important for the standard-siren analysis below, because once $c_T=1$ is enforced, the leading gravitational-wave effect is no longer a modified propagation speed but a modified amplitude damping driven by the evolution of the effective Planck mass~\cite{Belgacem:2017ihm, Belgacem:2018lbp, Lagos:2019kds}. The running effective Planck mass is therefore the direct bridge between the theory and the gravitational-wave observables studied in this work.

A covariant form of the Extended Cuscuton action is~\cite{Iyonaga:2018vnu, Iyonaga:2020bmm}
\begin{widetext}
\begin{equation}
S^{\ec}=\frac{1}{2}\int\dd^4x\sqrt{-g}\left[\,G_{4}^\ec(\phi) R+G_{2}^\ec(\phi,X)-G_{3}^\ec(\phi,X)\Box\phi\,\right],
\label{eq:EC_Action}
\end{equation}
\begin{align}
    G_2^\ec(\phi,X) &= -f_1(\phi) + f_2(\phi) \sqrt{2X}
    -\left[ 2f_{3,\phi}(\phi)+4f_{4,\phi\phi}(\phi)+\frac{3{f_3}^2(\phi)}{4f_4(\phi)}\right]X
    + \Big[ f_{3,\phi}(\phi)+2f_{4,\phi\phi}(\phi)\Big] X \ln X \, ,\\
    G_3^\ec(\phi,X) & =\frac{1}{2}\left[\, f_3(\phi)+2f_{4,\phi}(\phi)\,\right]
    \ln{X } \, ,
    \qquad\qquad G_4^\ec(\phi)  = f_4(\phi) \, .
\end{align}
\end{widetext}
Here, $X\equiv-\tfrac{1}{2}g^{\mu\nu}\partial_{\mu}\phi\partial_{\nu}\phi>0$ is the kinetic term associated with the timelike gradient of the scalar field, and $\Box\equiv g^{\mu\nu}\nabla_{\mu}\nabla_{\nu}$ is the d'Alembertian operator. The total action is $S = S^{\ec}+S^{\m}$, where $S^{\m}$ denotes the matter action and defines the standard stress-energy tensor through $T_{\mu\nu}\equiv-\tfrac{2}{\sqrt{-g}}\frac{\delta S^{\m}}{\delta g^{\mu\nu}}\,.$ Reduced Planck units are used throughout, with $c=\hslash=1$ and the bare gravitational coupling normalized as $8\pi G_N = 1$. We adopt the convention in which the scalar field has dimensions of mass${}^{-1}$, so that the kinetic term $X$ is dimensionless, with the dimensional factors absorbed into the Extended Cuscuton functions.

At the background level, we assume a spatially flat FLRW metric and a homogeneous time-dependent scalar field,
\begin{equation}
    \dd s^2 = -\dd t^2 +a^2(t)\left(\dd r^2+r^2\dd\Omega^2\right)\,,\quad 
    \phi=\phi(t)\,.
\end{equation}
Throughout, the branch choice $\dot{\phi}<0$ is imposed~\footnote{This convention is the opposite of the one used in~\cite{Iyonaga:2020bmm}, but it does not change the physical consequences of the model.}. This fixes the sign of $\sqrt{2X}$ and selects a future-directed scalar-field gradient.

Following Ref.~\cite{Iyonaga:2020bmm}, we restrict the analysis to the following polynomial subclass, which reduces the scalar-field equation of motion to a linear algebraic equation (see Eq.~\eqref{eq:scalar_constraint_dimful}),
\begin{align}
    f_4(\phi)&=c_{1}\,\phi^2+c_{2}\,\phi+1\,,
    \label{eq:f4_choice}\\
    f_3(\phi)&=0\,,
    \label{eq:f3_choice}\\
    f_2(\phi)&=c_{3}\,\phi+c_{4}\,,
    \label{eq:f2_choice}\\
    f_1(\phi)&=6c_{5}\,\phi^2+6c_{6}\,\phi+2\Lambda\,,
    \label{eq:f1_choice}
\end{align}
where $c_i$ are constant couplings and $\Lambda$ is the constant term in $f_1$. This is the simplest non-minimally coupled choice that keeps the background equations algebraically tractable (\textit{i.e.}, the equation of motion of the scalar field is analytically solvable) while retaining enough structure to allow departures from $\Lambda$CDM. The GR limit corresponds to $\phi = 0$ and $f_4 = 1$, while the locally measured Newton constant is $G_N^{\rm phys} = G_N/f_4(\phi)$.

For this subclass, the background equations take the form
\begin{align}
      &\phi\, \left(4 c_{5}-c_{3}\,H +4 c_{1}\,H^2 \right)+2 c_{6}- c_{4}\,H +2 c_{2}\,  H^2=0\,,
      \label{eq:scalar_constraint_dimful}\\
      &H^2=\frac{1}{3}\left(\rho+\Lambda \right)+ \phi^2 \left(c_5 -c_1\, H^2\right)+\phi \left(c_6 -c_2\,  H^2\right)\,,
      \label{eq:friedmann_dimful}   
\end{align}
where $H\equiv\dot a/a$ denotes the Hubble rate. The standard matter sector satisfies the continuity equation $\dot{\rho}+3H(\rho+P)=0\,,$ where $\rho$ and $P$ are the matter energy density and pressure. For the forecast analysis, the matter sector is taken to be a dust perfect fluid, namely $P=0$ and $\rho=\rho_0\,(a/a_0)^{-3}$, where $\rho_0$ and $a_0$ denote their present-day values.

Equation~\eqref{eq:scalar_constraint_dimful} can be solved algebraically for $\phi$, provided that $4 c_{5}-c_{3}\,H +4 c_{1}\,H^2\neq0$ throughout the cosmological evolution.

For a monotonically expanding universe, it is convenient to use the redshift $z:=\frac{a_0}{a}-1$ as the independent variable instead of cosmic time. We then introduce the dimensionless quantities
\begin{gather}
    \Omega_{\m,0}=\frac{\rho_0}{3H^2_0}\,,\quad    \Omega_{\Lambda}=\frac{\Lambda}{3H^2_0}\,,\quad     E=\frac{H}{H_0}\,, \quad    \vf=\phi\,H_0\,,\nn\\
    \tc_1=\frac{c_1}{H_0^2}\,,\quad
    \tc_2=\frac{c_2}{H_0}\,,\quad
    \tc_3=\frac{c_3}{H_0^3}\,,\nn\\
    \tc_4=\frac{c_4}{H_0^2}\,,\quad\tc_5=\frac{c_5}{H_0^4}\,,\quad\tc_6=\frac{c_6}{H_0^3}\, ,
\end{gather}
where $H_0\equiv H(z=0)$. In terms of these quantities,\footnote{Note that $\Omega_{\m,0}$ and $\Omega_\Lambda$ are defined here with respect to the bare coupling $8\pi G_N=1$. For comparison with a GR-based interpretation, in which the critical density is normalized using the local physical Newton constant $G_N^{\rm phys}=G_N/f_4(\vf_0)$, the relevant reference quantities are $\Omega_{\m,0}/f_4(\vf_0)$ and $\Omega_\Lambda/f_4(\vf_0)$.} the background dynamics is given by
\begin{align}
    E^2&=\Omega_{\m,0}(1+z)^3+\Omega_{\Lambda}+\Omega_{\ec}\,,
    \label{eq:Friedmann}\\
    \Omega_{\ec}&=\vf^2 \left(\tc_5 -\tc_1\, E^2\right)+\vf \left(\tc_6 -\tc_2\,  E^2\right)\,,
    \label{eq:Omc}\\
    \vf&=-\frac{2\tc_6-\tc_4\,E+2\tc_2\,E^2}{4\tc_5-\tc_3\,E+4\tc_1\,E^2}\,.
    \label{eq:vf}
\end{align}
Equation~\eqref{eq:Friedmann} is implicit in $E(z)$ and must therefore, in general, be solved numerically at each redshift.

The normalization condition $E(0)=1$ gives
\begin{equation}
    \Omega_{\m,0} = 1 - \Omega_{\Lambda} - \Omega_{\ec}\big|_{z=0}\, ,
    \label{eq:Om0}
\end{equation}
with
\begin{equation}
    \Omega_{\ec}\big|_{z=0} = \vf_0^2\left(\tc_5-\tc_1\right)+\vf_0\left(\tc_6-\tc_2\right)\,,
    \label{eq:Omc0}
\end{equation}
where
\begin{equation}
    \vf_0 = -\frac{2\tc_6-\tc_4+2\tc_2}{4\tc_5-\tc_3+4\tc_1}\,.
    \label{eq:vf0}
\end{equation}
Thus $\Omega_{\m,0}$ is not an independent parameter, but is determined by $(H_0,\,\Omega_\Lambda,\,\tc_1,\ldots,\tc_6)$.

Differentiating Eq.~\eqref{eq:vf} and using Eq.~\eqref{eq:Friedmann}, one obtains the following first-order system:
\begin{widetext}
\begin{align}
    \vf'(z)&=2\Pi_\ec E'(z)\,,
    \label{eq:varphi_prime} \qquad \Pi_\ec(z)=\frac{2\tc_4\tc_5 - \tc_3\tc_6 +8E\!\left(\tc_1\tc_6-\tc_2\tc_5 \right) + \left(\tc_2\tc_3 - 2\tc_1\tc_4\right)E^2}{\left(4\tc_5 - \tc_3\, E + 4\tc_1\, E^2\right)^2}\,,\\
    E'(z)& = \frac{3\left(E^2 -\Omega_\Lambda- \Omega_{\ec}\right)}{(1+z)\,E\left[2 + 2\vf\!\left(\tc_2 + \tc_1\,\vf\right) - \Pi_\ec\!\left(\tc_4 + \tc_3\,\vf - 4E\!\left(\tc_2 + 2\tc_1\,\vf\right)\right)\right]}\,.
    \label{eq:E_prime}
\end{align}
\end{widetext}
The quantity $2\Pi_\ec$ is the derivative of the algebraic scalar solution $\vf$ with respect to the expansion rate $E$. It therefore controls how $\vf'(z)$ and $E'(z)$ are coupled in the reduced first-order system.

A key observable for the forecast analysis is the gravitational-wave luminosity distance. In luminal scalar-tensor theories, the tensor propagation equation can still differ from its GR form through an additional friction term even when $c_T=1$. Integrating this modified damping leads to a gravitational-wave luminosity distance different from the electromagnetic one~\cite{Belgacem:2017ihm, Belgacem:2018lbp}. When the effect is due to a running effective Planck mass, one can write~\cite{Lagos:2019kds}
\begin{equation}
    \frac{d_L^{\rm GW}(z)}{d_L^{\rm em}(z)}=\frac{M_*(0)}{M_*(z)}\,.
    \label{eq:dgw_dem_planck}
\end{equation}
Here $M_*(z)$ is the effective Planck mass entering the tensor sector. In the present theory, $M_*^2 \propto G_4 = f_4\,$ so that, within the homogeneous cosmological approximation adopted here,
\begin{align}
    d_L^{\rm em}(z) &= (1+z)\,\frac{1}{H_0}
        \int_0^z\frac{\dd z'}{E(z')}\,,
        \label{eq:dL_em}\\
    d_L^{\rm GW}(z) &=d_L^{\rm em}(z)\sqrt{\frac{f_4(\vf_0)}{f_4(\vf(z))}}\,.
    \label{eq:dL_gw}
\end{align}
Equation~\eqref{eq:dL_gw} is therefore not an ad hoc parametrization, but the direct consequence of the running non-minimal coupling in the tensor sector. Here, local values of $f_4$ at the source and observer are identified with the corresponding background FLRW values $f_4(\vf(z))$ and $f_4(\vf_0)$, neglecting any departure from the homogeneous solution due to local perturbations or screening effects in the vicinity of the source or the observer.

\subsection{Viability conditions}

Having specified the analytically tractable subclass, we now impose the theoretical and observational conditions that define the viable parameter space. At the background level, the expansion rate must remain real and positive, $E(z)>0$, and the matter abundance must also be positive, $\Omega_{\m,0}>0$.

The algebraic solution for the scalar field must stay well defined, which requires the denominator of Eq.~\eqref{eq:vf} never to vanish:
\begin{equation}
    4\tc_5 - \tc_3\,E + 4\tc_1\,E^2 \neq 0\,.
    \label{eq:viability_varphi_denominator}
\end{equation}
Since the tensor-sector gravitational coupling is controlled by $f_4$, the corresponding effective Planck mass squared must remain positive,
\begin{equation}
    f_4(\vf)=\tc_1\,\vf^2 + \tc_2\,\vf + 1 > 0\,.
    \label{eq:viability_f4_positive}
\end{equation}
The branch choice must also be preserved throughout the cosmological evolution. Using $\dd z/\dd t=-(1+z)H$, this is equivalent to
\begin{equation}
    -\infty<\dot{\phi}<0
    \quad \Longleftrightarrow \quad
    0<\vf'(z)<\infty\,.
    \label{eq:viability_branch}
\end{equation}
Finally, the background solution is required to describe an accelerating universe today,
\begin{equation}
    0<-\frac{\dot H_0}{H_0^2}<1
    \quad \Longleftrightarrow \quad
    0<E'(z=0)<1\,.
    \label{eq:viability_acceleration}
\end{equation}

Since the goal is to study genuinely non-minimal Extended Cuscuton models, the analysis is restricted to $\tc_1\neq0$. In the polynomial subclass considered here, this permits the high-redshift background to recover the matter-dominated scaling $H\propto a^{-3/2}$, in analogy with~\cite{Iyonaga:2020bmm}. Moreover, an overall rescaling of the scalar field can be used to fix the absolute value of $\tc_1$, so that without loss of generality $ \tc_1=\pm1\,$.
The sign remains physical because $c_1$ multiplies $\phi^2$ in the non-minimal coupling.

We then impose observational priors associated with the time variation of the gravitational coupling. A time-dependent non-minimal coupling implies a time-dependent local Newton constant. In the Extended Cuscuton dark-energy realization, one has $G_N\propto1/f_4$ up to normalization conventions. The Lunar Laser Ranging bound on $\dot G_N/G_N$ therefore translates into a direct bound on the background solution. The current $2\sigma$ constraint is imposed as $|{\dot{G}_N}/{G_N}|_{z=0}<\varepsilon_{\rm LLR}$ with $\varepsilon_{\rm LLR}=1.52\times10^{-13}\,{\rm yr}^{-1}$~\cite{Williams:2004qba, Hofmann:2018myc}. In terms of the dimensionless variables, this becomes
\begin{equation}
    H_0 \left|\frac{(\tc_2 + 2\tc_1\vf_0)\,\vf'(z=0)}{1 + \tc_2\vf_0 + \tc_1\vf_0^2}\right| < \varepsilon_{\rm LLR}\,.
    \label{eq:LLR}
\end{equation}
Here $H_0$ is converted to ${\rm yr}^{-1}$ through
$ H_0\,[{\rm yr}^{-1}] = H_0\,[{\rm km\,s^{-1}\,Mpc^{-1}}]\times 1.022\times10^{-12}$.

A Big Bang Nucleosynthesis prior must also be imposed, since primordial element formation constrains the gravitational coupling at early times through the expansion rate. The relevant quantity is
\begin{equation}
    \frac{G_{N,{\rm BBN}}}{G_{N,0}} = \frac{f_4(\vf_0)}{f_4(\vf_\infty)}\,,
    \label{eq:BBN_ratio}
\end{equation}
where $\vf_\infty=-\tc_2/(2\tc_1)$ denotes the asymptotic high-redshift value of the scalar field, obtained from Eq.~\eqref{eq:vf} in the limit $z\to\infty$ for $\tc_1\neq0$. The $2\sigma$ bound~\cite{Alvey:2019ctk,Pitrou:2018cgg,Fields:2019pfx} is then imposed as
\begin{equation}
    \left|\frac{G_{N,{\rm BBN}}}{G_{N,0}} - 1\right| =
    \left|\frac{f_4(\vf_0)}{f_4(\vf_\infty)} - 1\right| < 0.06\,.
    \label{eq:BBN}
\end{equation}

\subsection{Simplest submodels with non-degenerate \texorpdfstring{$\Lambda$}{Lambda}CDM limit}\label{subsec:ec_submodels}

We now impose additional restrictions that define the benchmark submodels analyzed in this work. The guiding principle is that the model should admit a direct and non-degenerate $\Lambda$CDM limit. In the analytically tractable subclass, this requirement is non-trivial because the general scalar solution Eq.~\eqref{eq:vf} can approach zero in two qualitatively different ways: either the numerator tends to zero, or the denominator becomes large. The first possibility corresponds to a direct approach to the $\Lambda$CDM limit through small departures in the couplings. The second instead suppresses the background deviation through large absolute values of $\tc_3$ and $\tc_5$, since a large denominator in Eq.~\eqref{eq:vf} makes $\vf$ small and correspondingly reduces the departure from $\Lambda$CDM.

In the first route, the numerator of Eq.~\eqref{eq:vf} is driven to zero by small values of $\tc_2$, $\tc_4$, and $\tc_6$, which forces $\vf\to 0$; since the non-minimal coupling is $f_4 = \tc_1\vf^2+\tc_2\vf+1$, this in turn implies $f_4\to 1$, recovering the standard gravitational coupling of General Relativity. This is the non-degenerate route: the $\Lambda$CDM limit is approached through the genuine vanishing of the non-minimal gravitational coupling deviation, $f_4(\vf)\to 1$, and of the cuscuton coupling $f_2(\vf)$; the parameters $\tc_2$, $\tc_4$, and $\tc_6$ are in principle constrained by the data for any finite prior range.
In the second route, $\vf\to 0$ is instead achieved by making the denominator $4\tc_5 - \tc_3\,E + 4\tc_1\,E^2$ large through growing absolute values of $|\tc_3|$ and $|\tc_5|$, while the numerator remains finite $\mathcal{O}(1)$.
The physical mechanism here is qualitatively different. The coupling $\tc_5$ multiplies $\phi^2$ in $f_1(\phi)$ and therefore acts as a mass-like term for the Extended Cuscuton scalar in the potential sector of the action: a large (positive) $\tc_5$ suppresses $\vf$ by driving the scalar to the minimum of its own effective potential rather than by removing the non-minimal coupling from the theory. Similarly, $\tc_3$ is the field-dependent coefficient of $f_2(\vf) = \tc_3\vf + \tc_4$, which modulates the amplitude of the classical cuscuton contribution $f_2(\vf)\sqrt{2X}$; a large $|\tc_3|$ suppresses $\vf$ through this field-dependent cuscuton coupling.

The second route is potentially problematic from an inference perspective: it leads to a prior-dominated posterior under Bayesian sampling. 
Along the direction $|\tc_3|,|\tc_5|\to\infty$, the scalar solution is suppressed by the denominator of Eq.~\eqref{eq:vf}, and the background observables can approach the $\Lambda$CDM limit while remaining only weakly sensitive to the precise values of these parameters. In a Bayesian analysis with broad uniform priors, this may lead to prior-dominated posterior support along poorly constrained directions, rather than to constraints driven by the likelihood. Similar prior-volume effects are known to occur in cosmological extensions of $\Lambda$CDM with degenerate limiting directions~\cite{Gomez-Valent:2022hkb, Holm:2023laa, Herold:2024enb, Colgain:2023bge}. A dedicated treatment of the full $\tilde c_3,\tilde c_5$ sector would therefore require methods designed to diagnose such directions, such as profile-likelihood analyses or suitable reparametrizations~\cite{Herold:2024enb, Holm:2023laa, Paradiso:2024yqh}. We leave this broader analysis to future work and focus here on the non-degenerate benchmark sector defined by $\tilde c_3=\tilde c_5=0$.

In the benchmark models considered here, we remove the degenerate route entirely at the level of the model definition by imposing 
\begin{equation}
    \tc_1=\pm1\,,
    \qquad
    \tc_3=\tc_5=0\,.
    \label{eq:simplest_subclass}
\end{equation}
Then, the background equations of the model reduce to:
\begin{align}
    E^2&=\Omega_{\m,0}(1+z)^3+\Omega_{\Lambda}+\Omega_{\ec}\,,
    \label{eq:SM_Friedmann}\\
    \Omega_{\ec}&= -\tc_1\,\vf^2 E^2+\vf \left(\tc_6 -\tc_2\,  E^2\right)\,,
    \label{eq:SM_Omc}\\
    \vf(E)&=-\frac{2\tc_6-\tc_4\,E+2\tc_2\,E^2}{4\tc_1\,E^2}\,.
    \label{eq:SM_vf}
\end{align}

Under this restriction, the only path to the $\Lambda$CDM limit is the non-degenerate one: $\vf\to 0$ requires $\tc_2\to 0$, $\tc_4\to 0$, and $\tc_6\to 0$, which simultaneously drives $f_4\to 1$ and removes the Extended Cuscuton contribution from both the background expansion and the tensor sector. The posterior structure is then free from the degenerate direction, and the parameters have a direct physical interpretation as measures of the departure from General Relativity through the non-minimal coupling.
This isolates the simplest non-minimally coupled Extended Cuscuton subclass compatible with a matter-dominated past and a luminal tensor sector, while keeping the approach to the $\Lambda$CDM limit controlled by parameters with a direct physical meaning.

We then impose a further restriction that fixes the late-time de Sitter limit. The algebraic structure of the system already guarantees the existence of a de Sitter attractor~\cite{Iyonaga:2020bmm}: once $E\to E_{\rm dS}={\rm const}<1$, Eqs.~\eqref{eq:varphi_prime} and~\eqref{eq:E_prime} force $\vf\to\vf_{\rm dS}={\rm const}$. In general, however, the asymptotic expansion rate is $E_{\rm dS}^2 = \Omega_\Lambda + \Omega_{\rm ec}^{\rm dS}$, with $\Omega_{\rm ec}^{\rm dS}\neq0$ unless an additional condition is imposed on the couplings. To reduce the parameter-space dimension further and identify $\Lambda$ directly with the asymptotic de Sitter value, we require
\begin{equation}
    \Omega_{\ec}\to0
    \qquad \text{for} \qquad z\to-1\,,
    \label{eq:de_sitter_limit_Omc}
\end{equation}
which is a restriction on the selected submodels rather than a generic property of the analytically tractable Extended Cuscuton subclass. Physically, this condition ensures that $\Lambda$ plays the role of the cosmological constant of the asymptotic de Sitter attractor exclusively, with no residual late-time contribution from the Extended Cuscuton sector. In this way, the $\Lambda$CDM limit is recovered exactly when the remaining non-minimal couplings vanish. It also reduces the effective dimension of parameter space by one. The late-time de Sitter attractor is then fixed by
\begin{equation}
    E^2\to\Omega_\Lambda
    \qquad \text{for} \qquad z\to-1\,.
    \label{eq:de_sitter_limit_E}
\end{equation}
The corresponding asymptotic scalar value is
\begin{equation}
    \vf_{\dS}=-\frac{2\tc_6-\tc_4\,\sqrt{\Omega_\Lambda}+2\tc_2\,\Omega_\Lambda}{4\tc_1\,\Omega_\Lambda}\,.
    \label{eq:SM_vf_ds}
\end{equation}

Evaluating Eq.~\eqref{eq:SM_Omc} at the asymptotic de Sitter point $E^2=\Omega_\Lambda$, using Eq.~\eqref{eq:SM_vf_ds}, one obtains a factorized expression for $\Omega_{\ec}^{\dS}$,
\begin{equation}
    \Omega_{\ec}^{\dS}=
    -\frac{\left(2\tc_6-\tc_4\sqrt{\Omega_\Lambda}+2\tc_2\Omega_\Lambda\right)
    \left(6\tc_6-\tc_4\sqrt{\Omega_\Lambda}-2\tc_2\Omega_\Lambda\right)}{16\tc_1\Omega_\Lambda}\,.
    \label{eq:Omc_ds_factorized}
\end{equation}
Imposing $\Omega_{\ec}^{\dS}=0$ then selects two branches,
\begin{align}
    \tc_6&=\frac{1}{2}\tc_4\sqrt{\Omega_\Lambda}-\tc_2\Omega_\Lambda\,,
    \label{eq:tc6_branch1}\\
    \tc_6&=\frac{1}{6}\tc_4\sqrt{\Omega_\Lambda}+\frac{1}{3}\tc_2\Omega_\Lambda\,.
    \label{eq:tc6_branch2}
\end{align}
The first branch corresponds to an asymptotic scalar that tends to zero, $\vf_{\dS}=0$. The second corresponds to a non-vanishing asymptotic scalar for which the Extended Cuscuton contribution still vanishes asymptotically. Combining these two branches with the two possible signs of $\tc_1$ yields the four benchmark submodels considered in this work.
\begin{table}[!ht]
\caption{Extended Cuscuton benchmark submodels considered in this work. A convenient independent parameter set is $(H_0,\,\Omega_\Lambda,\,\tilde c_2,\,\tilde c_4)$, with $\tilde c_6$ fixed by the de Sitter condition and $\Omega_{m,0}$ derived from Eq.~\eqref{eq:Om0}.\vspace{3pt}}
\label{tab:submodels}
\centering
\renewcommand{\arraystretch}{2.5}
\setlength{\tabcolsep}{8pt}
\begin{tabular}{clc}
\toprule\\[-31pt]
\bf Submodel (SM) & \bf Asymptotic condition & $\boldsymbol{\vf_{\dS}}$ \\
\midrule
SM01:\quad $\tc_1=+1$  & $\tc_6=\dfrac{1}{2}\tc_4\sqrt{\Omega_\Lambda}-\tc_2\Omega_\Lambda$ & $0$ \\
SM02:\quad $\tc_1=+1$  & $\tc_6=\dfrac{1}{6}\tc_4\sqrt{\Omega_\Lambda}+\dfrac{1}{3}\tc_2\Omega_\Lambda$ & $\neq 0$ \\[5pt]
\arrayrulecolor{gray!50}\hline\arrayrulecolor{black}\\[-25pt]
SM03:\quad $\tc_1=-1$  & $\tc_6=\dfrac{1}{2}\tc_4\sqrt{\Omega_\Lambda}-\tc_2\Omega_\Lambda$ & $0$ \\
SM04:\quad $\tc_1=-1$  & $\tc_6=\dfrac{1}{6}\tc_4\sqrt{\Omega_\Lambda}+\dfrac{1}{3}\tc_2\Omega_\Lambda$ & $\neq 0$ \\[3pt]
\bottomrule
\end{tabular}
\end{table}

Notice that, when early-time datasets are included, the radiation contribution must be restored in the background equation,
\begin{equation}
    E^2=\Omega_{\rm r,0}(1+z)^4+\Omega_{\m,0}(1+z)^3+\Omega_\Lambda+\Omega_{\ec}\,.
    \label{eq:Friedmann_radiation}
\end{equation}
At the perturbation level, one must also distinguish carefully between three different quantities: the tensor-sector coupling $f_4$, which controls $d_L^{\rm GW}$; the locally measured Newton constant $G_N$, constrained by Eq.~\eqref{eq:LLR}; and the effective scalar-sector coupling $G_{\rm eff}$ entering the growth of matter perturbations~\cite{Iyonaga:2020bmm}. These quantities are related, but they are not observationally interchangeable.

The phenomenological analysis can then be organized around this four-submodel basis. The same non-minimal coupling $f_4$ controls the background evolution, the local-gravity priors, and the gravitational-wave signal, which makes it possible to trace a single physical ingredient across the different observational sectors considered in the following sections.

\section{Current cosmological data and data analysis} \label{sec:realdata}
\begin{table*}[t]
\centering
\caption{Parameter priors adopted in the analysis. Flat priors are imposed as listed below. In addition, BBN and LLR viability conditions introduce effective hard prior cuts within the likelihood evaluation. A Gaussian prior is applied to the supernova absolute magnitude $M$.}
\label{tab:priors}

\begin{tabular}{l c l}
\hline
Parameter & Prior & Motivation \\
\hline

$\Omega_\mathrm{\Lambda}$ & $[0.1,\,0.9]$ 
& Broad range consistent with dark-energy domination. \\

$H_0$ & $[50,\,90]$ 
& Covers both Planck and SH0ES measurements with conservative margin. \\

$\tc_2$ & $[-1.9,\,1.9]$ 
& Strict subset of $|c_2|<2$ required for $f_4>0$ when $c_1=+1$, chosen slightly inside the boundary \\ & & to avoid numerical instabilities. \\

$\tc_4$ & $[-5,\,5]$ 
& Weakly constrained by current cosmological data; broad prior adopted. \\

$M$ (flat) & $[-20,\,-18]$ 
& Conservative range for the SN absolute magnitude. \\

$M$ (Gaussian) & $\mathcal{N}(-19.214,\,0.037^2)$ 
& Gaussian prior with mean $-19.214$ and standard deviation $0.037$. \\

\hline
\end{tabular}
\end{table*}

We now describe the cosmological dataset used to carry out a Monte Carlo Markov Chain (MCMC) analysis to test the background evolution of the Cuscuton model given in equations~\eqref{eq:SM_Friedmann}--\eqref{eq:SM_vf}.

\subsection{Cosmological Dataset}

\begin{itemize}
    \item \textbf{Cosmic Chronometers:} We make use of 32 cosmic chronometer data points covering the redshift interval $0.07 \leq z \leq 1.965$, assembled from several sources~\citep{Moresco:2022phi, Moresco:2024wmr}. Derived through the differential age technique applied to massive galaxies undergoing passive evolution, these data yield direct measurements of $H(z)$ that are free from any reliance on distance ladder calibrations or assumptions about the underlying cosmological model. The complete dataset can likewise be accessed via GitLab \footnote{\url{https://gitlab.com/mmoresco/CCcovariance}}.
    \item \textbf{Supernovae Type-Ia (SN):} Our analysis draws on the extensive ``Pantheon Plus Sample'' of Type-Ia Supernovae (SN-Ia), which contains 1701 supernovae distributed across the redshift interval 0.01 to 2.26~\citep{Scolnic}. This dataset folds in the SH0ES distance anchors, relying on host cepheid galaxies for its calibration~\citep{sntable}. To calibrate the SNe, we fixed the absolute magnitude at $M_{B}$ = $-19.214 \pm 0.037$ magnitudes. This value emerges from combining geometric distance determinations based on Detached Eclipsing Binaries in the Large Magellanic Cloud (LMC)~\citep{piet}, the MASER NGC4258~\citep{reid}, together with recent parallax measurements of 75 Milky Way Cepheids obtained from Hubble Space Telescope (HST) photometry~\citep{riess21} and GAIA Early Data Release 3 (EDR3)~\citep{lind20a,lind20b}. This stands as the most precise and up-to-date model-independent estimate of the Absolute Magnitude currently available.\\
    Once the Absolute Magnitude of standard candles such as Type-Ia Supernovae (SNe-Ia) is known, their distance follows directly from their measured apparent magnitude or flux. The connection between the apparent magnitude of SNe-Ia and their relative distance modulus is given by:
    \begin{equation}\label{dl}
    {D}_{L}=10^{(\mu - 25)/5} \text{ Mpc}.
    \end{equation}
    In this expression, $\mu = m_b - M_B$ is the distance modulus, with $m_b$ being the apparent magnitude of SNe-Ia and $M_B$ the corresponding absolute magnitude.
    \item \textbf{Baryon Acoustic Oscillations (BAO):} For our baseline analysis, we adopt the DESI DR1 BAO measurements as provided through the official DESI likelihood \footnote{The DESI likelihood is publicly accessible at \url{https://github.com/cosmodesi/desilike}}. This compilation comprises the BGS sample over $0.1 < z < 0.4$, the LRG1 and LRG2 samples covering $0.4 < z < 0.6$ and $0.6 < z < 0.8$ respectively, the combined LRG3+ELG1 sample within $0.8 < z < 1.1$, the ELG2 sample over $1.1 < z < 1.6$, the quasar sample spanning $0.8 < z < 2.1$, and the Lyman-$\alpha$ Forest sample extending across $1.77 < z < 4.16$. We stress that the DESI BAO measurements adopted in this analysis are those of the first data release (DESI~DR1) and \emph{not} the more recent second release (DESI~DR2). This choice reflects the analysis pipeline that was already in place at the time of writing rather than any preference for the earlier dataset. We have verified that repeating the analysis with DR2 does not materially alter our findings: the central values are stable, and the principal effect is a modest tightening of the constraints, as expected from the larger DR2 sample. The improvement is therefore sub-dominant and not the main focus of this work. A full reanalysis with DR2 is straightforward and is the natural next step, but we do not attempt it here, since it would leave the main conclusions of the paper essentially unchanged.

\end{itemize}

\subsection{Parameter priors and \texorpdfstring{$\chi^2$}{chi2} minimisation}
The parameter space explored in this work is summarized in Table~\ref{tab:priors}. We impose broad, physically motivated flat priors on all cosmological parameters in order to avoid artificially restricting the posterior volume. The prior on $\Omega_\Lambda$ spans $[0.1,\,0.9]$, ensuring consistency with a dark-energy dominated late-time Universe while remaining agnostic about the precise value. For the Hubble constant, we adopt $H_0 \in [50,\,90]\,\mathrm{km\,s^{-1}\,Mpc^{-1}}$, a conservative range encompassing both Planck and SH0ES determinations with additional margin. 

The parameter $\tc_2$ is restricted to the interval $[-1.9,\,1.9]$, chosen as a strict subset of the theoretical bound $|\tc_2|<2$, which guarantees $f_4>0$ for $\tc_1=+1$ (\textit{i.e.}, SM01 and SM02).
Indeed, in the high-redshift limit, the scalar field approaches $\vf_\infty = -\tc_2/(2\tc_1)$, and therefore $f_4(\vf_\infty) = 1 - \tc_2^2/(4\tc_1)$.
For $\tc_1=+1$, positivity of $f_4(\vf_\infty)$ requires $|\tc_2|<2$. For $\tilde c_1=-1$, the same interval is kept for uniformity.
The slight inward shift from the exact boundary avoids numerical instabilities near the viability limit. For $\tc_4$, which remains weakly constrained by current data, we adopt a broad prior $[-5,\,5]$. The supernova absolute magnitude is assigned both a conservative flat prior $M \in [-20,\,-18]$ and a Gaussian prior $\mathcal{N}(-19.214,\,0.037^2)$ reflecting calibration constraints.

In addition to the explicit priors listed in Table~\ref{tab:priors}, theoretical viability conditions from Big Bang Nucleosynthesis (BBN) and Lunar Laser Ranging (LLR) are implemented directly at the likelihood level. These act effectively as hard prior cuts but are enforced numerically rather than imposed analytically.

Parameter estimation is performed via $\chi^2$ minimisation, defined as
\begin{equation}
    \chi^2(\theta) = \sum_i \frac{\big[D_i^{\mathrm{obs}} - D_i^{\mathrm{th}}(\theta)\big]^2}{\sigma_i^2},
\end{equation}
where $\theta$ denotes the set of model parameters, $D_i^{\mathrm{obs}}$ are the observational data points, $D_i^{\mathrm{th}}$ the corresponding theoretical predictions, and $\sigma_i$ the associated uncertainties. For datasets with non-diagonal covariance matrices, the quadratic form $\chi^2 = \Delta \mathbf{D}^T \mathbf{C}^{-1} \Delta \mathbf{D}$ is employed.

\section{Constraints from CC, BAO, and SN observations}\label{constraints}

\begin{figure*}
\begin{minipage}{0.48\textwidth}
    \includegraphics[width=\textwidth]{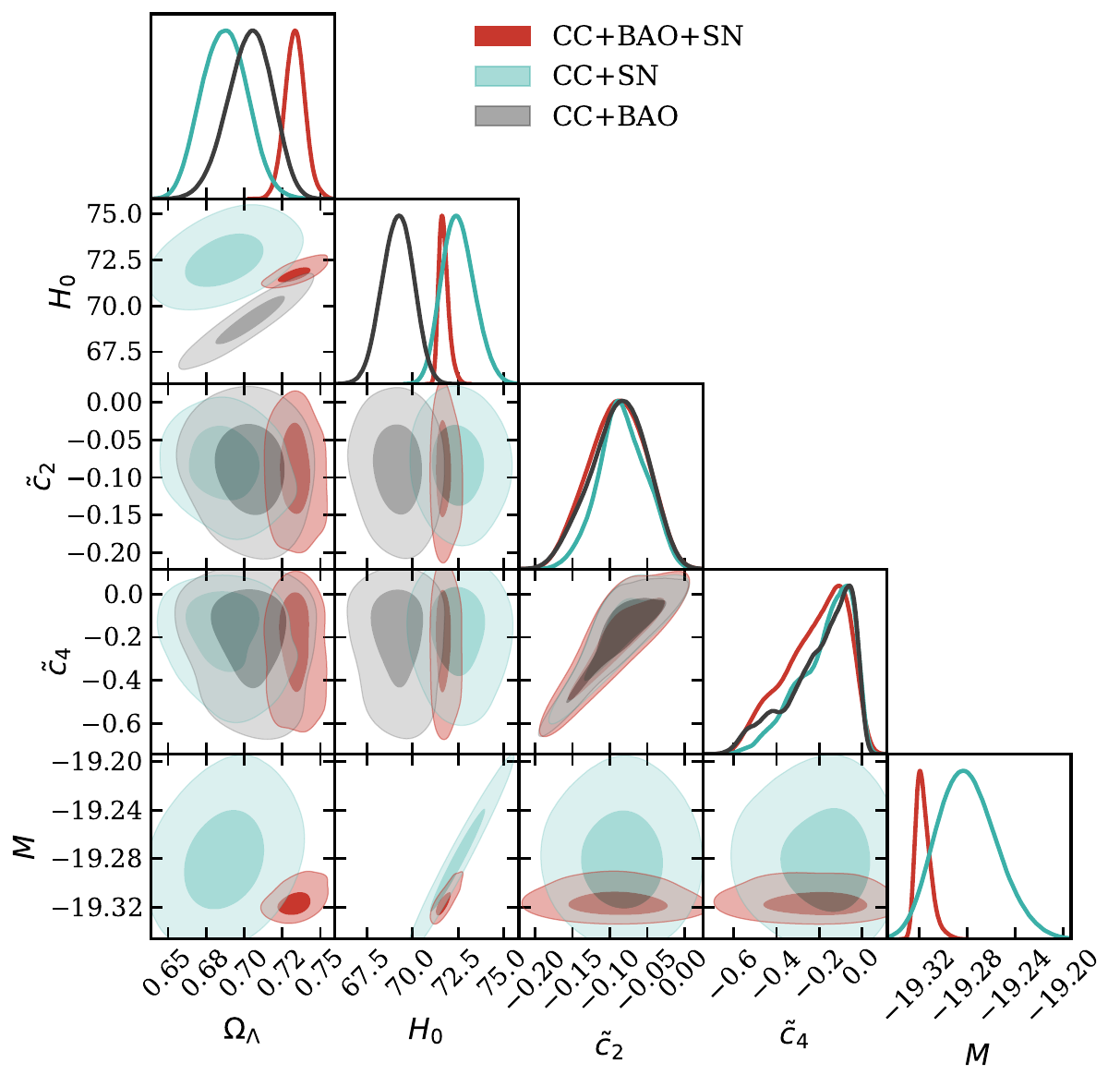}\\
    (a) \textbf{SM01}
\end{minipage}
\hfill
\begin{minipage}{0.48\textwidth}
    \includegraphics[width=\textwidth]{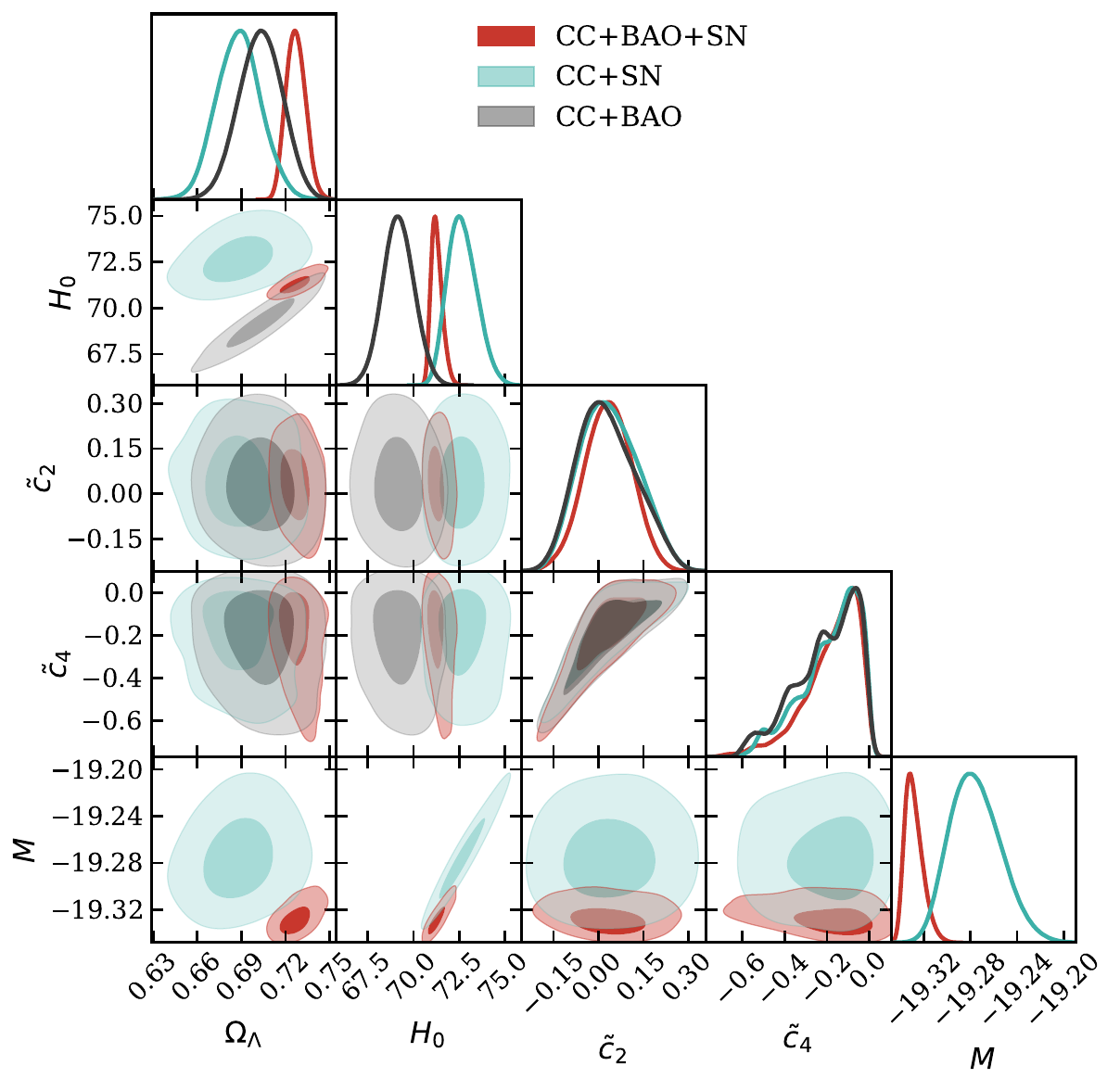}\\
    (b) \textbf{SM02}
\end{minipage}\\[8pt]
\begin{minipage}{0.48\textwidth}
    \includegraphics[width=\textwidth]{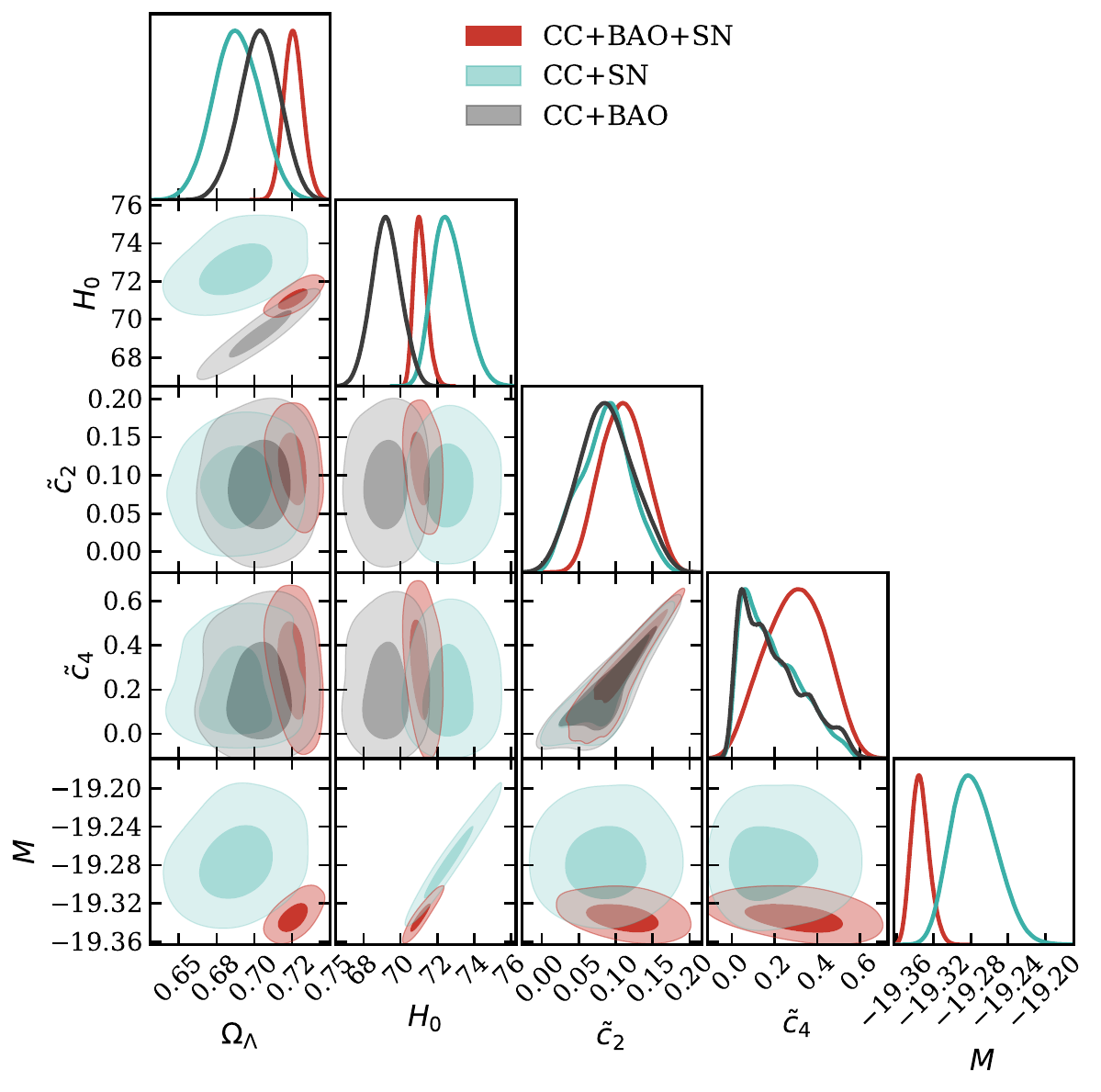}\\
    (c) \textbf{SM03}
\end{minipage}
\hfill
\begin{minipage}{0.48\textwidth}
    \includegraphics[width=\textwidth]{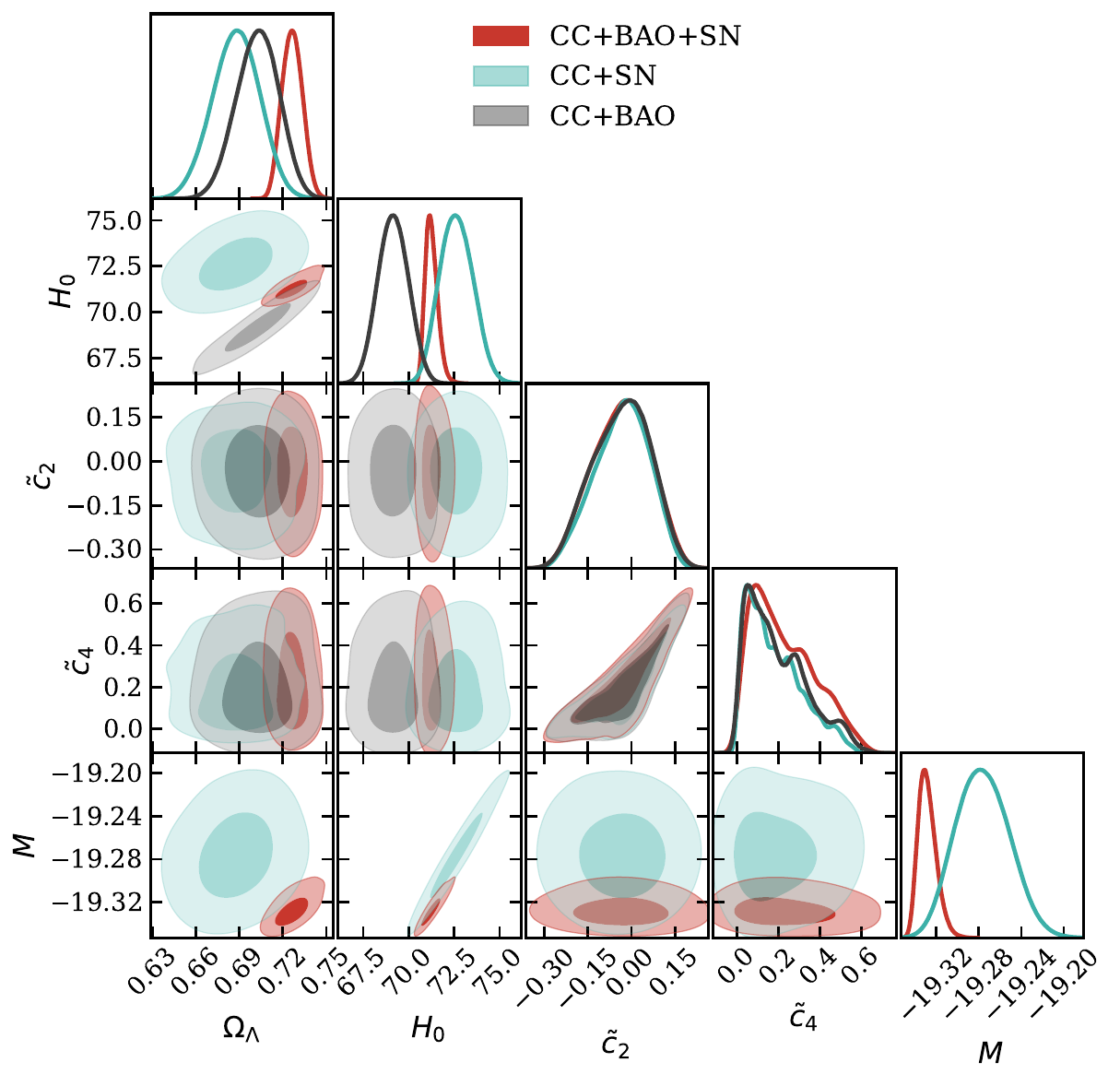}\\
    (d) \textbf{SM04}
\end{minipage}
\caption{Posterior distributions (68\% and 99\% credible intervals) of the sampled parameters $(H_0,\, \Omega_{\Lambda},\, \tc_2,\, \tc_4,\, M)$ for Extended Cuscuton submodels 1--4: (a) SM01, (b) SM02, (c) SM03, (d) SM04.}
\label{fig:corner_other_probes}
\end{figure*}

\begin{table*}[!ht]
\caption{Median values and 68\% credible intervals for the parameters of Models 1--4 ($H_0$ in km s$^{-1}$Mpc$^{-1}$) obtained from different data combinations.}
\renewcommand{\arraystretch}{1.5}
\label{tab:constraints}
\small
\setlength{\tabcolsep}{5pt}
\begin{tabular}{c|c|c c c c}
\hline
\textbf{Parameter\,} & \textbf{\,Data\,} & \textbf{SM01} & \textbf{SM02} & \textbf{SM03} & \textbf{SM04} \\
\hline
& CC+BAO+SN 
& $0.73^{+0.01}_{-0.01}$ & $0.73^{+0.01}_{-0.01}$ & $0.73^{+0.01}_{-0.01}$ & $0.73^{+0.01}_{-0.01}$ \\
{\boldmath$\Omega_{\Lambda}$} & CC+SN 
& $0.69^{+0.01}_{-0.01}$ & $0.69^{+0.01}_{-0.02}$ & $0.69^{+0.02}_{-0.02}$ & $0.69^{+0.02}_{-0.02}$ \\
& CC+BAO 
& $0.70^{+0.02}_{-0.01}$ & $0.70^{+0.01}_{-0.02}$ & $0.70^{+0.01}_{-0.01}$ & $0.70^{+0.02}_{-0.01}$ \\
\hline
& CC+BAO+SN 
& $71.71^{+0.20}_{-0.31}$ & $71.27^{+0.24}_{-0.35}$ & $71.09^{+0.30}_{-0.40}$ & $71.25^{+0.27}_{-0.39}$ \\
{\boldmath$H_0$} & CC+SN 
& $72.50^{+0.81}_{-0.96}$ & $72.66^{+0.71}_{-0.88}$ & $72.64^{+0.76}_{-0.97}$ & $72.63^{+0.92}_{-0.93}$ \\
& CC+BAO 
& $69.20^{+0.86}_{-0.87}$ & $69.19^{+0.86}_{-0.83}$ & $69.23^{+0.77}_{-0.81}$ & $69.16^{+0.81}_{-0.81}$ \\
\hline
& CC+BAO+SN 
& $-0.09^{+0.04}_{-0.04}$ & $0.03^{+0.08}_{-0.08}$ & $0.11^{+0.03}_{-0.03}$ & $-0.03^{+0.11}_{-0.10}$ \\
{\boldmath$\tc_2$} & CC+SN 
& $-0.08^{+0.03}_{-0.03}$ & $0.04^{+0.09}_{-0.11}$ & $0.09^{+0.04}_{-0.03}$ & $-0.03^{+0.11}_{-0.09}$ \\
& CC+BAO 
& $-0.09^{+0.04}_{-0.03}$ & $0.03^{+0.09}_{-0.11}$ & $0.09^{+0.04}_{-0.04}$ & $-0.03^{+0.11}_{-0.09}$ \\
\hline
& CC+BAO+SN 
& $-0.22^{+0.19}_{-0.10}$ & $-0.17^{+0.16}_{-0.06}$ & $0.30^{+0.15}_{-0.14}$ & $0.22^{+0.10}_{-0.20}$ \\
{\boldmath$\tc_4$} & CC+SN 
& $-0.17^{+0.16}_{-0.07}$ & $-0.18^{+0.17}_{-0.07}$ & $0.19^{+0.08}_{-0.18}$ & $0.18^{+0.08}_{-0.17}$ \\
& CC+BAO 
& $-0.20^{+0.19}_{-0.07}$ & $-0.20^{+0.19}_{-0.07}$ & $0.19^{+0.07}_{-0.19}$ & $0.20^{+0.11}_{-0.19}$ \\
\hline
& CC+BAO+SN 
& $-19.32^{+0.00}_{-0.01}$ & $-19.33^{+0.01}_{-0.01}$ & $-19.33^{+0.01}_{-0.01}$ & $-19.33^{+0.01}_{-0.01}$ \\
{\boldmath$M$} & CC+SN 
& $-19.28^{+0.02}_{-0.03}$ & $-19.28^{+0.02}_{-0.02}$ & $-19.28^{+0.02}_{-0.03}$ & $-19.28^{+0.02}_{-0.03}$ \\
& CC+BAO 
& --- & --- & --- & --- \\
\hline
\end{tabular}
\end{table*}

The median values and $68\%$ credible intervals for all submodel parameters are summarised in Table~\ref{tab:constraints}. We can focus on SM01, since the following considerations apply to all the submodels.
The full data combination CC$+$BAO$+$SN provides the tightest constraints, yielding $\Omega_\Lambda = 0.73_{-0.01}^{+0.01}$ and $H_0 = 71.71_{-0.31}^{+0.20}$~km\,s$^{-1}$\,Mpc$^{-1}$, together with the shape parameters $\tc_2 = -0.09_{-0.04}^{+0.04}$ and $\tc_4 = -0.22_{-0.10}^{+0.19}$. \\
The two-dataset combinations follow the trend anticipated above: CC$+$SN prefers a higher expansion rate, $H_0 = 72.50_{-0.96}^{+0.81}$, driven by the SN calibration, whereas CC$+$BAO settles at a lower value, $H_0 = 69.20_{-0.87}^{+0.86}$, reflecting the CC$+r_d$ anchoring; this same shift is mirrored in $\Omega_\Lambda$, which moves from $0.69_{-0.01}^{+0.01}$ (CC$+$SN) to $0.70_{-0.01}^{+0.02}$ (CC$+$BAO).\\
The shape parameters $\tc_2$ and $\tc_4$ remain remarkably stable across all combinations, indicating that they are constrained by the overall shape of the expansion history rather than by its absolute calibration. Finally, the supernova absolute magnitude is tightly pinned at $M = -19.32_{-0.01}^{+0.00}$ in the full combination and $M = -19.28_{-0.03}^{+0.02}$ for CC$+$SN, while it is absent from the CC$+$BAO analysis, which does not include supernovae.\\

We observe a mild tension in $H_0$ and $\Omega_{\Lambda}$ between the data combinations CC$+$SN and CC$+$BAO. The reason is that cosmic chronometers alone constrain $H_0$ to lie around $67$--$68$~km\,s$^{-1}$\,Mpc$^{-1}$ in $\Lambda$CDM (see Table~2 of Ref.~\cite{Ruchika:2025mkx}), whereas BAO does not measure $H_0$ on its own: it only constrains the combination $H_0 r_d$~\cite{Evslin:2017qdn,Ruchika:2024lgi,Dutta:2019pio}. Consequently, in the CC$+$BAO combination, the $H_0$ of BAO is effectively calibrated by the $H_0$ inferred from CC, settling at the lower value $H_0 = 69.20_{-0.87}^{+0.86}$. In the CC$+$SN combination, on the other hand, $H_0$ is set by the supernovae through the absolute magnitude $M_B$ (coming from SH0ES)\cite{Ruchika:2024ymt,Ruchika:2023ugh}, which prefers a higher value, $H_0 = 72.50_{-0.96}^{+0.81}$.\\
In addition, we fix the sound horizon $r_d$ to its BBN-based value ($147.5$~Mpc). Since BAO constrains the combination $H_0  r_d$~\cite{Evslin:2017qdn}, this choice, together with the CC calibration, drives $H_0$ toward the lower end of our results. Fixing $r_d$ is well motivated: being the comoving distance travelled by the acoustic fluid up to the drag epoch, it is set entirely by early-universe physics prior to that epoch and is therefore independent of the late-time expansion we aim to reconstruct. The same behaviour can be seen in the corner plot (Fig.~\ref{fig:corner_other_probes}): the red contours denote CC$+$BAO$+$SN, the green CC$+$SN, and the grey CC$+$BAO. The offset between the grey and green contours in the $H_0$--$\Omega_\Lambda$ plane directly illustrates the mild tension discussed above.

Although the discussion above refers to SM01, the same qualitative and quantitative behaviour is shared by all four extended Cuscuton models: the constraints on $\Omega_\Lambda$, $H_0$ and $M$ agree across SM01--SM04 to well within $1\sigma$ for every data combination, with the largest shift in $H_0$ ($\Delta H_0 \simeq 0.6$~km\,s$^{-1}$\, Mpc$^{-1}$ between Models 1 and 3 for CC$+$BAO$+$SN) remaining below the $\sim1.5\sigma$ level.
The only model-dependent quantities are the shape parameters $\tc_2$ and $\tc_4$, which are expected since they encode the specific form of the Cuscuton contribution in each submodel. In particular, $\tc_2$ is negative at the $\sim2\sigma$ level for SM01 ($\tc_2 = -0.09^{+0.04}_{-0.04}$), positive at a similar significance for SM03 ($\tc_2 = 0.11^{+0.03}_{-0.03}$), and consistent with zero for SM02 and SM04, while $\tc_4$ changes sign between SM01--SM02 ($\tc_4 < 0$) and SM03--SM04 ($\tc_4 > 0$). The fact that $\tc_2$ and $\tc_4$ absorb the model dependence while leaving the background parameters essentially unchanged indicates that the four parametrisations differ mainly in how the departures from $\Lambda$CDM are encoded, rather than in the background expansion history they reconstruct from CC, BAO, and SN data.

\section{Gravitational wave data and data analysis}\label{sec:mockdata}
\subsection{Generation of Mock Gravitational Wave Catalogs}

To evaluate the scientific reach of 3G GW detectors, we construct synthetic binary neutron star (BNS) catalogs following the established methodology in~\cite{Califano:2022cmo, Califano:2022syd, Califano:2023fbq,Califano:2024xzt}. The redshift distribution of the GW sources is governed by the probability density function~\cite{Regimbau:2012ir,Cai:2016sby}:
\begin{equation}\label{rate:unit_of_redshift}
p(z) = \mathcal{N}\frac{R_m (z)}{1+z}\frac{dV(z)}{dz},
\end{equation}
where $\mathcal{N}$ denotes the normalization constant and $dV(z)/dz$ represents the comoving volume element. The source-frame merger rate per unit volume, $R_m(z)$, is defined by the convolution of the star formation rate (SFR) and the time delay distribution $P(t_d)$~\cite{Regimbau:2009rk,Regimbau:2016ike,Meacher:2015rex}:
\begin{equation}\label{merger_rate}
R_{m} (z) = R_{\rm m,0} \int_{t_{min}}^{t_{max}} R_f[t(z)-t_d] P(t_d) d t_d.
\end{equation}
In accordance with population synthesis results~\cite{Lipunov:1995ct,deFreitasPacheco:2005ub,Belczynski:2006br}, we assume $P(t_d)\propto t_d^{-1}$, adopting a minimum delay $t_{min}=20$ Myr and a maximum extending to the Hubble time~\cite{Meacher:2015iua}. The SFR $R_f(z)$ follows the Madau-Dickinson cosmic evolution~\cite{Madau:2014bja}:
\begin{equation}
R_f(z)=\left[1+(1+z_p)^{-\gamma-\kappa}\right]\frac{(1+z)^{\gamma}}{1+\left(\frac{1+z}{1+z_p}\right)^{\gamma+\kappa}},
\end{equation}
with parameters $\gamma=2.6$, $\kappa=3.1$, and $z_p=2$. We normalize the local merger rate to $R_m(z=0)=105.5_{-83.9}^{+190.2}$  Gpc$^{-3}$ yr$^{-1}$ as per recent LVK results~\cite{KAGRA:2021duu}.

The fiducial luminosity distance $d_L^{fid}(z)$ is computed assuming a flat $\Lambda$CDM cosmology with parameters derived from the \textit{Planck} 2018 legacy release: $H_0=67.66$ km s$^{-1}Mpc^{-1}$ and $\Omega_{m,0}=0.3111$~\cite{Planck:2018vyg}. The total number of observable mergers $N$ over an observation period $T_{obs}$ with a duty cycle $\mathcal{D} =0.85$ is given by:
\begin{equation}
N = T_{obs}\ \mathcal{D} \int_{0}^{10} \frac{R_m (z)}{1+z}\frac{dV(z)}{dz}dz.
\end{equation}

Individual source parameters $\Theta=\{m_1, m_2, d_L,\theta_{jn}, t_c, \phi_c, \psi, \rm{RA}, \rm{Dec}\}$ are sampled assuming isotropic sky localization and uniform orientations. Following LVK mass distributions for BNS, component masses are drawn uniformly in the range $[1,2.5] M_\odot$~\cite{KAGRA:2021duu}. We utilize the \texttt{GWFISH} software package~\cite{Dupletsa:2022scg} to compute the Signal-to-Noise Ratio (SNR) and the Fisher Information Matrix (FIM), retaining only events with SNR $>9$. The FIM $F_{ab}$ is defined as the inner product of the waveform derivatives:
\begin{equation}
\mathcal{F}_{ab} =\left(\frac{\partial h}{\partial \Theta^a}\middle| \frac{\partial h}{\partial \Theta^b}\right)_{\mathbf{\Theta} = \bar{\mathbf{\Theta}}}, 
\end{equation}
where $h(t,\mathbf{\Theta})$ is the GW strain, generated with the \texttt{IMRPhenomD\_NRTidalv2} waveform~\cite{Colleoni:2023ple} whose choice is based on the comparison to other waveforms in~\cite{Dupletsa:2022scg}.
The statistical uncertainty for the luminosity distance, $\sigma_{inst}$, is extracted from the inverse FIM:
\begin{equation}\label{eq:fim}
    \sigma_{\Theta^a}=\sqrt{(\mathcal{F}^{-1})^{ab}}\,.
\end{equation}
To represent realistic observations, we define the total uncertainty $\sigma_{d_L}$ by incorporating instrumental noise, weak lensing, and peculiar velocity corrections~\cite{Speri:2020hwc, Cen:1998nj}:
\begin{equation}\label{sigma_dl}
\sigma_{d_L}^2=\sigma_{inst}^2+\sigma_{lens}^2 +\sigma_{pec}^2,
\end{equation}
where
\begin{equation}
    \sigma_{lens}=0.066\left(\frac{1-(1+z)^{-0.25}}{0.25}\right)^{1.8}d_L(z)F(z),
    \end{equation}
and 
\begin{equation}
\sigma_{pec}=\left[ 1+\frac{c(1+z)^2}{H(z)d_L (z)}\right]\frac{\sqrt{\langle v^2\rangle}}{c}d_L (z)\,.
\end{equation}
Here, $F(z)= 1- \frac{0.3}{\pi /2}\arctan{\frac{z}{z_*}}$, with $z_*=0.073$~\cite{Speri:2020hwc}, and the averaged peculiar velocity $\sqrt{\langle v^2\rangle}$ is set to  $500$ km/s, in agreement with the observed values in galaxy catalogs~\cite{Cen:1998nj}. 
The specific detector coordinates and frequency ranges employed in our multi-detector network configurations (Einstein Telescope and Cosmic Explorer) are summarized in Table~\ref{tab:detectors}. 

\begin{table}[ht!]
\centering
\caption{Configurations and operational parameters for the 3G detector network~\cite{Borhanian:2020ypi}. ET components are co-located in Italy; CE-ID and CE-NM are located in Idaho and New Mexico, USA, respectively.}
\label{tab:detectors}
\begin{tabular}{l c c c c c }
\hline
Detector & Latitude & Longitude & \shortstack{x-arm\ azim.} & \shortstack{y-arm\ azim.} & $f_{ini}$ [Hz]\\
\hline
ET-1 & 0.7615 &  0.1833 & 0.3392 &5.5752& 2\\
    ET-2 & 0.7629 & 0.1841 & 4.5280 & 3.4808 & 2 \\
    ET-3 & 0.7627 & 0.1819 & 2.4336& 1.3864 & 2\\
    CE-ID & 0.7649 & -1.9692 &1.5708 & 0 & 5 \\
    CE-NM &0.5787 & -1.8584 &2.3562 & 0.7854& 5 \\
\hline
\end{tabular}
\end{table}

Our statistical analysis will employ three detector network configurations\footnote{In the ET+CE configuration, we consider the CE located in Idaho with 40 km arm length.}:
ET, ET + CE, ET + 2CE. The Table~\ref{tab: detected event} lists the number of detected GW events after one year of observations in the case of four sky-localization uncertainty thresholds, namely 10, 40, 100, and 1000 deg$^2$ at the 90\% confidence level, and distinguish BNS mergers with a generic orientation or a viewing angle\footnote{The viewing angle $\theta_v$ is the $\min(i, 180^{\circ} - i)$.} $\theta_v < 15^{\circ}$. Let us remark that only a small fraction of those events are expected to produce detectable high-energy emissions powered by the GRB relativistic jet, which is assumed to be perpendicular to the orbital plane.
\begin{table*}
    \centering
    \setlength{\tabcolsep}{0.4em}
        \caption{Number of events detected by ET, ET+CE and ET+2CE after one year of observations. We list the GW events with sky localization better than 10, 40, 100, and 1000 deg$^2$, and distinguish BNS mergers with a generic orientation or a viewing angle $\theta_v < 15^{\circ}$.}\label{tab: detected event}
    \begin{tabular}{c|c|c|c||c|c|c}
    \hline
    $\theta_v$& \multicolumn{3}{c||}{Any $\theta_v$}&\multicolumn{3}{c}{$\theta_v<15^{\circ}$}\\
    \hline
        Network & ET & ET+CE  & ET+2CE & ET & ET+CE  & ET+2CE \\
        \hline
        $N_{\rm det}$ &24374& 46516 &88358 & 2131 & 3563 & 4766 \\
        $N_{\rm det}(\Delta\Omega < 1000\, \text{deg}^2)$ &755  & 42668 & 85910& 135 & 3272 & 4685 \\
        $N_{\rm det}(\Delta\Omega < 100\, \text{deg}^2)$ & 139 & 19896 &39066 & 19 &  1940 &3233 \\
        $N_{\rm det}(\Delta\Omega < 40\, \text{deg}^2)$ &  46& 6178 &14470 & 4 & 699 & 1492\\
        $N_{\rm det}(\Delta\Omega < 10\, \text{deg}^2)$ &  4& 803 & 2012& 0 & 83 & 222 \\
        \hline
    \end{tabular}
\end{table*}

\subsection{Modeling Multi-Messenger Emission}\label{subsec: EM counterpart}

Following the framework established in~\cite{Ronchini:2022gwk,Branchesi:2023mws}, we evaluate the multi-messenger potential of 3G GW detectors by simulating synergistic observations across the electromagnetic spectrum. Our analysis focuses on three primary EM signatures: the prompt $\gamma$-ray emission from Gamma-Ray Bursts (GRBs), the subsequent X-ray afterglows, and the optical transients associated with kilonovae (KNe).

\subsubsection{GRB Prompt Emission and Jet Structure}\label{subsec: prompt emission}

We assume that a detectable relativistic jet is produced in 20\% of BNS mergers. To characterize the emission, we adopt a structured jet model based on the properties of GRB 170817A, assuming a universal angular distribution for the local emissivity $\epsilon(\theta_v)$ and the bulk Lorentz factor $\Gamma(\theta_v)$~\cite{Postnov:1999gj,Salafia:2022dkz}:
\begin{align}
\label{eq:epsilon} \epsilon(\theta_v) &= \frac{4\pi \epsilon_c}{1+\left( \frac{\theta_v}{\theta_c}\right)^{s_{\epsilon}}},\\
\label{eq:Gamma} \Gamma(\theta_v) &= 1 + \frac{\Gamma_0 - 1 }{1+\left( \frac{\theta_v}{\theta_c}\right)^{s_{\Gamma}}}.
\end{align}
In alignment with~\cite{Ronchini:2022gwk}, we utilize the parameters $s_\epsilon=s_\Gamma=4$, a core angle $\theta_c=3.4^\circ$, an initial Lorentz factor $\Gamma_0 = 500$, and a peak emissivity $4\pi\epsilon_c=3\times10^{53}$ erg.

The detectability of these events depends on the received photon flux $\mathcal{F}$:
\begin{equation}\label{eq:flux}
\mathcal{F} = \frac{L_{\rm iso}}{4\pi d_L^2} \times k(z), \quad L_{\rm iso} = \frac{2 E_{\rm iso}}{\langle t_{GRB}\rangle},
\end{equation}
where ⟨$t_{GRB}$⟩=22 s. $\mathcal{F}$ incorporates the isotropic equivalent luminosity $L_{\rm iso}$ and a cosmological $k$-correction~\cite{Salafia:2015vla} which accounts for the redshift of the photon spectrum $N(E)$, modeled here by a Band function~\cite{Band:1993eg,Nava:2010ig}. The peak energy $E_p$ is sampled from a log-normal distribution with $\log_{10}(\mu_E/{\rm keV})=3.2$ and $\sigma_E=0.36$. For an off-axis observer at angle $\theta_v$, the isotropic equivalent energy $E_{\rm iso}$ is determined by the Doppler-boosted integration of the jet profile:
\begin{equation}\label{eq:E_iso}
E_{\rm iso}(\theta_v) = \int \frac{\zeta^3(\theta,\phi,\theta_v)}{\Gamma(\theta)}\epsilon(\theta)d\Omega,
\end{equation}
where the Doppler factor $\zeta$ is defined by the velocity $\beta(\theta)$ and the geometric angle $\alpha$ between the velocity vector and the line of sight:
\begin{equation}
\zeta(\theta,\phi,\theta_v)=\frac{1}{\Gamma(\theta)[1 - \beta(\theta)\cos(\alpha(\theta,\phi,\theta_v))]}.
\end{equation}

To quantify joint GW-EM detections, we simulate the performance of the \textit{THESEUS}-XGIS instrument. We assume an 85\% duty cycle and a Field of View (FOV) of $\sim 2$ sr ($P_{\rm det} \approx 0.16$) within the 2-150 keV band. An event is classified as a joint detection if its flux exceeds the XGIS threshold of $3\times10^{-8}$erg cm$^{-2}$s$^{-1}$~\cite{THESEUS:2021uox}. The resulting counts for the ET, ET+CE, and ET+2CE networks are presented in Table~\ref{tab:number-events-combined}, showing close agreement with previous literature~\cite{Ronchini:2022gwk, Branchesi:2023mws, Califano:2024xzt}.

\subsubsection{X-ray Afterglow Emission}\label{subsec: afterglow}

The temporal evolution of the GRB afterglow is simulated using the \texttt{afterglowpy} Python package~\cite{Ryan:2019fhz}. The resulting light curves are determined by the jet geometry (\ref{eq:epsilon},~\ref{eq:Gamma}) and a set of microphysical parameters describing the shock physics and the circumburst environment. These include the interstellar medium (ISM) number density $n_0$, the electron energy distribution power-law index $p$, and the energy fractions partitioned into the magnetic field ($\epsilon_B$) and electrons ($\epsilon_e$).

Following the methodology in~\cite{Ronchini:2022gwk}, we fix $p=2.2$, $\epsilon_e=0.1$, and the jet wing extension $\theta_w=15^\circ$. To account for environmental diversity, we sample the remaining parameters from the intervals identified in~\cite{Fong:2015oha}: $\epsilon_B\in[0.01,0.1]$, the radiative efficiency $\eta\in[0.01,0.1]$, and $n_0\in [3,15]\times10^{-3}$ cm$^{-3}$.

Detection prospects are evaluated for the \textit{THESEUS} Soft X-ray Imager (SXI), characterized by a $0.5$ sr field of view and an arcmin-scale localization precision. We adopt a flux threshold of $1.8\times10^{-11}$erg cm$^{-2}$s$^{-1}$ in the $0.3-5$ keV band~\cite{THESEUS:2021uox}. Furthermore, we assess the synergy between SXI and XGIS, the latter providing a broader 2 sr coverage at higher energies ($>$2 keV). The predicted detection counts are summarized in Table~\ref{tab:number-events-combined}. We note that our estimates are slightly more conservative than those in~\cite{Ronchini:2022gwk, Branchesi:2023mws}, as \texttt{afterglowpy} does not currently account for high-latitude emission contributions~\cite{Ascenzi:2020kxz}.

\subsubsection{Kilonova Emission and Optical Follow-up}\label{subsec: KN}

The quasi-isotropic optical transient, or kilonova (KN), provides a unique electromagnetic signature that is independent of the binary inclination. We model the KN luminosity using the \texttt{redback} pipeline~\cite{Sarin:2023khf}, implementing the analytical framework from~\cite{Metzger:2016pju, Villar:2017oya}.

For the optical follow-up, we consider the Vera C. Rubin Observatory (VRO), utilizing its 8.4-meter aperture and 9.6 deg$^2$ field of view~\cite{LSST:2008ijt}. Given the high sensitivity of VRO, we adopt a Target of Opportunity (ToO) strategy~\cite{Andreoni:2021epw,Cowperthwaite:2018gmx,LSST:2018bbx}, selecting only BNS events with GW sky-localization uncertainties $\Delta\Omega<40$ deg$^2$~\cite{Branchesi:2023mws}. A joint detection is defined by a 5$\sigma$ significance in both $g$ and $z$ (or $g$ and $i$) filters during the first two nights post-merger, assuming 180-second exposures ($m_g^{lim}\sim26$, $m_z^{lim}\sim24.4$). This dual-filter approach is essential for distinguishing the KN color evolution from contaminating transients~\cite{LSSTTransient:2018dvm}.

Based on an allocation of 3600 survey hours per year, the VRO could potentially monitor up to $\sim$1200 mergers~\cite{Alfradique:2022tox}. Our projected joint GW+KN detection rates for various 3G network configurations are presented in Table~\ref{tab:number-events-combined}.

\begin{table}[h!]
\centering
\caption{Projected joint GW and EM detections for ET, ET+CE, and ET+2CE networks over 1, 5, and 10-year missions. The table distinguishes between prompt $\gamma$-ray (XGIS), X-ray afterglow (SXI and SXI+XGIS), and optical KN (VRO) counterparts.}
\label{tab:number-events-combined}
\small
 \begin{tabular}{c|c|c|c|c}
    \hline
     \multicolumn{5}{c}{\bf Prompt Emission}\\
            \hline
        Instrument&Years & ET & ET+CE & ET+2CE\\
        \hline
        \multirow{3}{*}{\makecell{THESEUS-XGIS\ ($\gamma$-ray)}} &1&10&25  &31 \\
       
        &5& 40 & 89 &176 \\
        
        &10& 99  &185& 327\\
        \hline
    \hline 
    \multicolumn{5}{c}{\bf Afterglow}\\
    \hline
    Instrument & Years & ET & ET+CE & ET+2CE\\
    \multirow{3}{*}{\makecell{THESEUS\\SXI+XGIS\\ (X-ray)}} &1&11& 12  &16 \\
    &5& 37 & 51 &87 \\
    &10&71  & 98& 135\\
    \hline
    \hline
      \multicolumn{5}{c}{\bf Kilonovae}\\
    \hline
       Instrument & Years & ET & ET+CE & ET+2CE\\
        \hline
        \multirow{3}{*}{\makecell{VRO\ (Optical)}} 
        &1&21&603  &{{765}} \\
        &5& 79 & 2973 &3743 \\
        &10&161  &{{5938}} & 7495\\
        \hline
    \end{tabular}
\end{table}

\section{Cosmological Parameter Estimation}\label{sec:results}
To quantify the precision with which 3G detector networks can constrain the expansion history of the Universe, we perform a Bayesian inference analysis using nested sampling. Our reference model for the mock catalogs is the flat $\Lambda$CDM model, hereafter referred to as the \emph{fiducial} model. As discussed in Sec.~\ref{sec:ecm}, we examine four submodels of the non-dynamical dark energy scenarios in Horndeski gravity and assess our ability to recover the fiducial $\Lambda$CDM baseline from these more complex frameworks.

The sampling is carried out with the \texttt{nessai} nested sampler~\cite{Williams:2021qyt}, accessed through the \texttt{bilby} inference framework~\cite{Ashton:2018jfp}. The \texttt{nessai} algorithm uses normalizing flows to construct efficient proposals within nested sampling, which is advantageous for parameter spaces with non-trivial posterior structure and for likelihoods that are computationally expensive to evaluate.

For a set of $N$ independent GW observations with associated electromagnetic counterparts, the global likelihood $p(\mathbf{d}|\bm{\lambda})$ is the product of individual event likelihoods:
\begin{equation}
p(\mathbf{d}|\bm{\lambda}) = \prod_{i=1}^{N} p(d_i|\bm{\lambda}),
\end{equation}
where $\bm{\lambda}$ represents the vector of cosmological parameters. In the standard $\Lambda$CDM case, $\bm{\lambda}=\{H_0,\Omega_\Lambda\}$. For the four scenarios in Horndeski gravity, the parameter space is enlarged to $\bm{\lambda}=\{H_0,\Omega_{\Lambda},\tilde{c}_2,\tilde{c}_4\}$, while is derived from the closure condition~\eqref{eq:Om0}, and $\tilde c_6$ is fixed by the corresponding asymptotic de Sitter condition~\eqref{eq:tc6_branch1} and~\eqref{eq:tc6_branch2}. The remaining couplings are fixed by the submodel definition, namely $\tilde c_1=\pm1$ and $\tilde c_3=\tilde c_5=0$. 

The single-event likelihood for a source at redshift $z$ is assumed to be Gaussian:
\begin{equation}
p(d_i|\bm{\lambda}) \propto \exp{\left(-\frac{1}{2}\frac{(d_i - d_L(z,\bm{\lambda}))^2}{\sigma^2_{d_L}}\right)},
\end{equation}
where $d_L(z,\bm{\lambda})$ is computed from the background expansion rate $E(z)$ of the corresponding cosmological model, Eqs.~\eqref{eq:SM_Friedmann}--\eqref{eq:SM_vf}, which reduces to $\Lambda$CDM when $\tilde{c}_2=\tilde{c}_4=0$.

Following Bayesian inference, the posterior distribution $p(\bm{\lambda}|\bm{d})$ is proportional to the product of the likelihood and the prior distributions $\pi(\bm{\lambda})$:
\begin{equation}
p(\bm{\lambda}|\mathbf{d}) \propto p(\mathbf{d}|\bm{\lambda})\,\pi (\bm{\lambda}).
\end{equation}
We adopt uninformative flat priors for all parameters: $H_0\in U(50,90)$, $\Omega_\Lambda\in U(0,1)$, $\tilde{c}_2\in U(-2,2)$ and $\tilde{c}_4\in U(-5,0)$ in case $\tilde{c}_1=1$ and $\tilde{c}_4\in U(0,5)$ in case $\tilde{c}_1=-1$. The latter are branch-consistency conditions rather than an observational prior, because the monotonicity of the scalar field implies $\tc_4\lessgtr0$ for $\tc_1=\pm1$.

The analysis considers three observational scenarios and three observational networks: five years of joint detections for Prompt Emission (PE, via THESEUS-XGIS), Afterglows (AF, via THESEUS-SXI+XGIS), and Kilonovae (KN, via VRO).  

\subsection{Fitting the fiducial model}

\begin{widetext}
\begin{figure*}
\centering
{\includegraphics[width=0.32\textwidth]{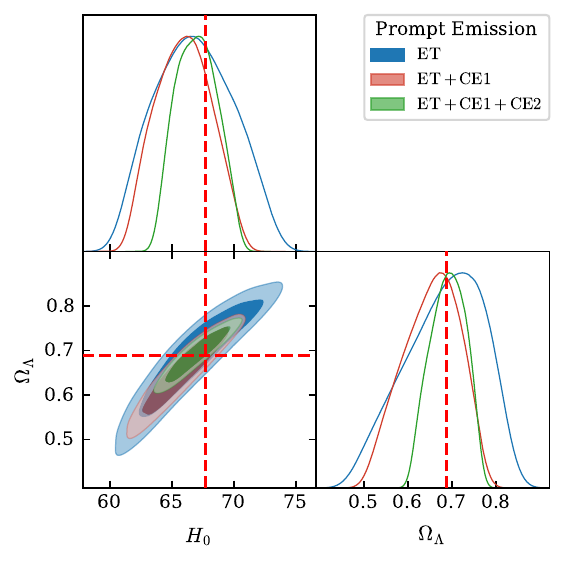}}
{\includegraphics[width=0.32\textwidth]{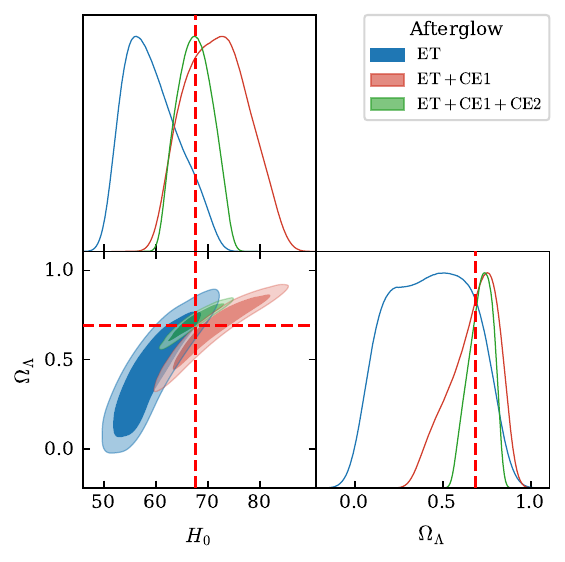}}
{\includegraphics[width=0.32\textwidth]{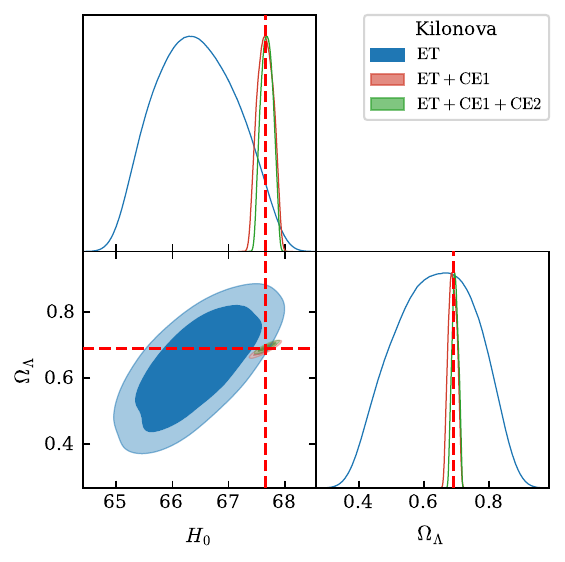}}
\caption{Posterior distributions (68\% and 95\% credible intervals) of the sampled parameters $(H_0,\, \Omega_{\Lambda})$ for the $\Lambda$CDM model in the three EM counterpart channels: blue, red, and green contours correspond to the ET, ET+CE, and ET+2CE detector networks, respectively. Red dashed lines mark the fiducial $\Lambda$CDM values ($H_0=67.66$, $\Omega_\Lambda=0.6889$).}
\label{fig:corner_mock_LCDM}
\end{figure*}
\end{widetext}

To check whether the mock catalogues were correctly built without introducing any bias, neither in the statistical analysis nor in the construction of the catalogues, we fit the fiducial flat $\Lambda$CDM model to the mock data.
The inferred median values and 68\% credible intervals for the fiducial $\Lambda$CDM parameters are summarized in Table~\ref{tab: result LCDM}. They show that the input cosmology is consistently recovered within the inferred uncertainties, thereby validating both the construction of the mock catalogues and the statistical inference pipeline. The plots in Fig~\ref{fig:corner_mock_LCDM} show the posterior distributions with their 68\% and 95\% credible intervals, for the sampled parameters $(H_0,\, \Omega_\Lambda)$ in the three electromagnetic counterpart channels. The blue, red, and green contours correspond to the ET, ET+CE, and ET+2CE detector networks, respectively, while the red dashed lines mark the fiducial $\Lambda$CDM values, $H_0=67.66$ and $\Omega_{\Lambda}=0.6889$.

\begin{table}[!ht]
\centering
\caption{Median values and 68\% credible intervals for the $\Lambda$CDM parameters ($H_0$ in km s$^{-1}$Mpc$^{-1}$) recovered from the PE, AF, and KN mock catalogs across three 3G network configurations.}
\renewcommand{\arraystretch}{1.5}
\label{tab: result LCDM}
\small
\begin{tabular} { c |c| c c c}
\hline
\textbf{Parameter\,} & \textbf{\,Network\,} & \textbf{\,PE (5 yr)} & \textbf{AF (5 yr)} & \textbf{KN (5 yr)} \\
\hline 
& ET 
& $66.14^{+4.39}_{-4.68}$ 
& $60.23^{+8.86}_{-6.04}$ 
& $66.34^{+1.00}_{-0.99}$ \\
{\boldmath$H_0$} & ET+ CE 
& $65.73^{+3.05}_{-3.36}$ 
& $70.17^{+7.91}_{-8.48}$ 
& $67.65^{+0.19}_{-0.19}$ \\
& ET+2CE 
& $66.84^{+2.30}_{-2.37}$ 
& $66.92^{+4.61}_{-4.82}$ 
& $67.68^{+0.15}_{-0.15}$ \\
\hline
& ET 
& $0.67^{+0.11}_{-0.17}$ 
& $0.49^{+0.29}_{-0.30}$ 
& $0.62^{+0.16}_{-0.18}$ \\
{\boldmath$\Omega_{\Lambda}$} & ET+ CE 
& $0.65^{+0.08}_{-0.11}$ 
& $0.67^{+0.20}_{-0.27}$ 
& $0.69^{+0.02}_{-0.02}$ \\
& ET+2CE 
& $0.69^{+0.05}_{-0.06}$ 
& $0.71^{+0.09}_{-0.12}$ 
& $0.70^{+0.01}_{-0.01}$ \\
\hline
\end{tabular}
\end{table}

\section{Forecasts with 3G Gravitational Wave detectors}\label{forecast}

We now forecast the precision with which next-generation GW detector networks can constrain the cosmological parameters of the Extended Cuscuton submodels described in Sec.~\ref{sec:ecm}. To this end, we carry out a Bayesian analysis using 3G GW mock catalogues for the ET, ET+CE, and ET+2CE detector networks. The median values and 68\% credible intervals are reported in Tables~\ref{tab:ec_sm01_sm02_5yr} and~\ref{tab:ec_sm03_sm04_5yr} for the selected benchmark submodels SM01--SM02 and SM03--SM04, respectively.

\begin{figure*}
\centering
{\includegraphics[width=0.336\textwidth]{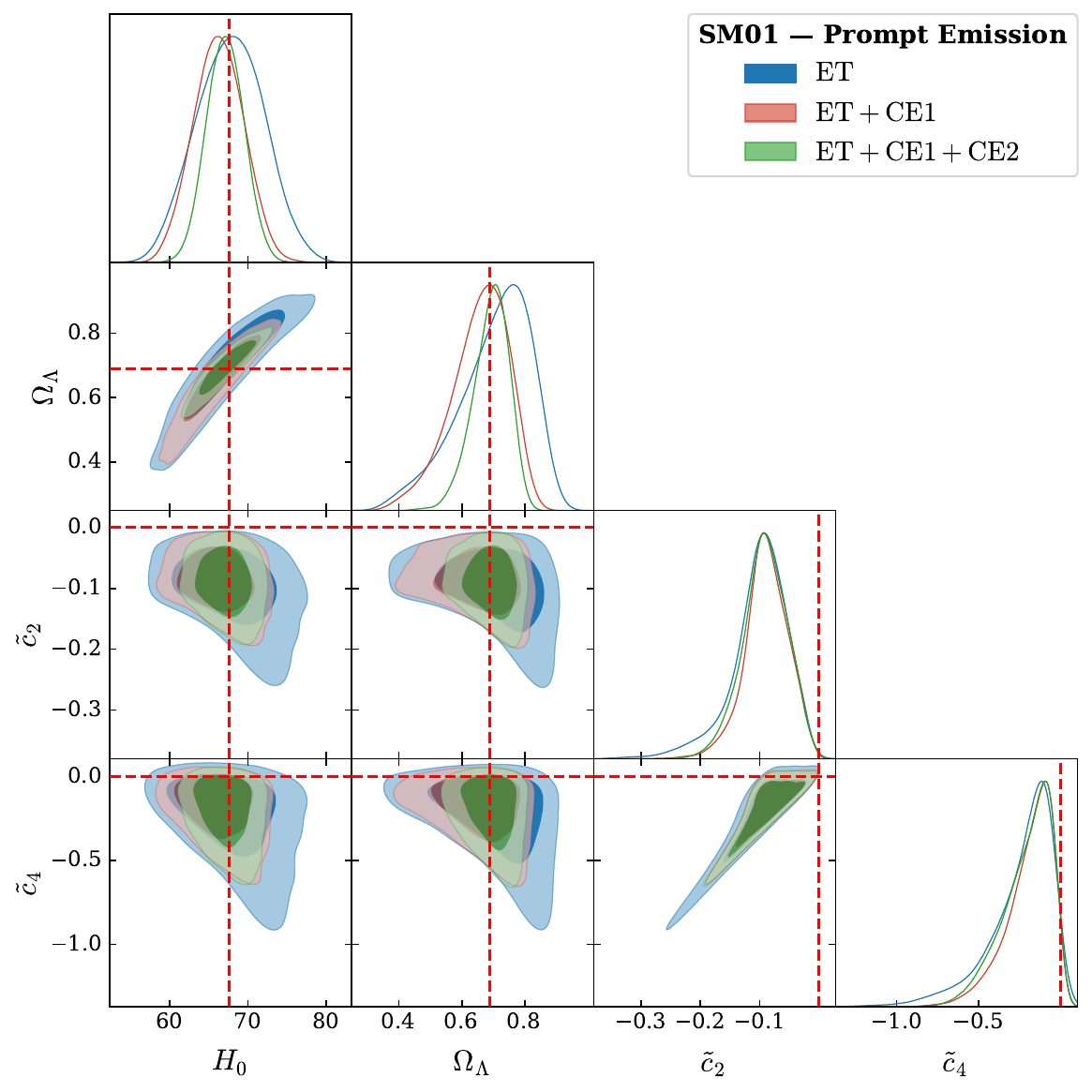}}\hspace{-5pt}
{\includegraphics[width=0.336\textwidth]{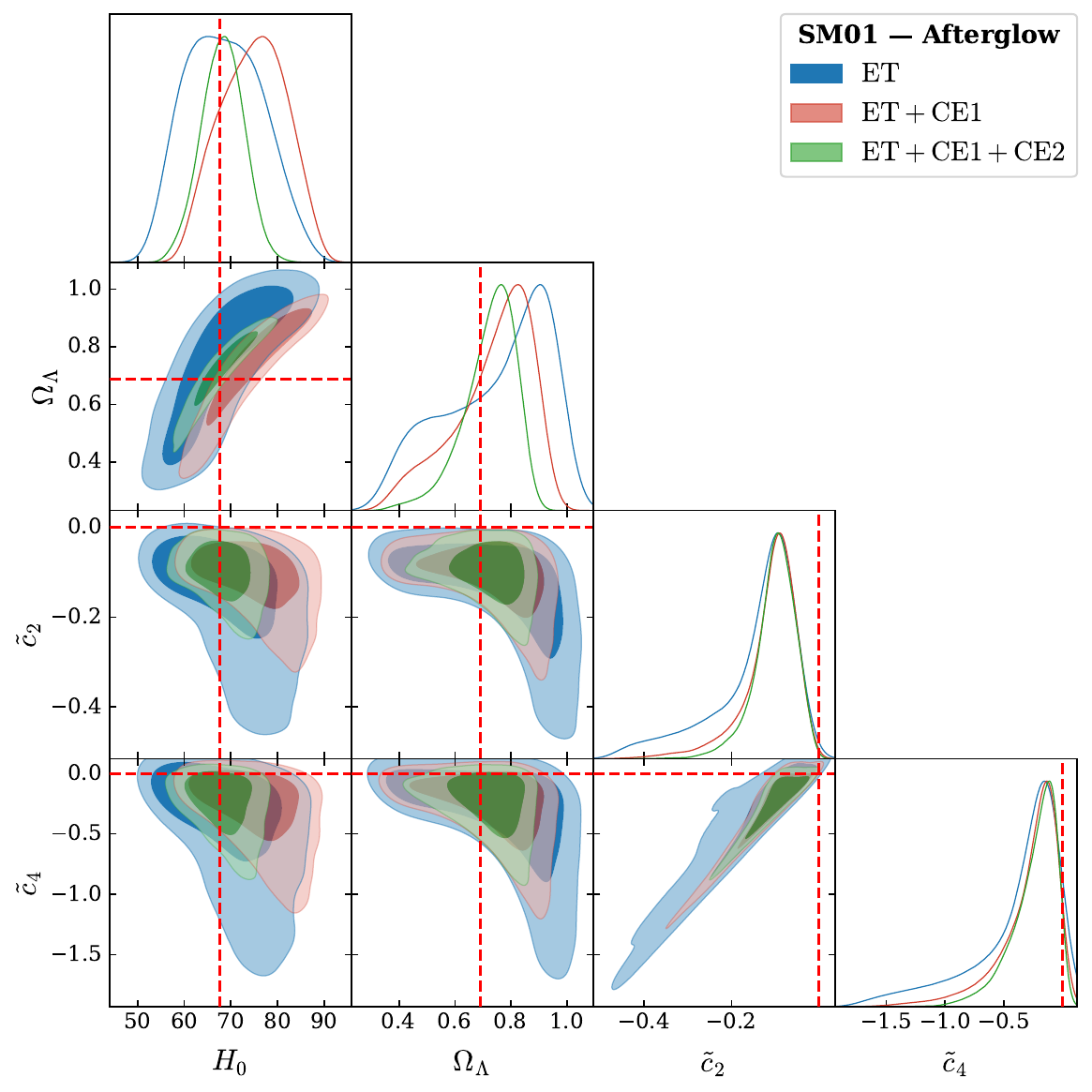}}\hspace{-5pt}
{\includegraphics[width=0.336\textwidth]{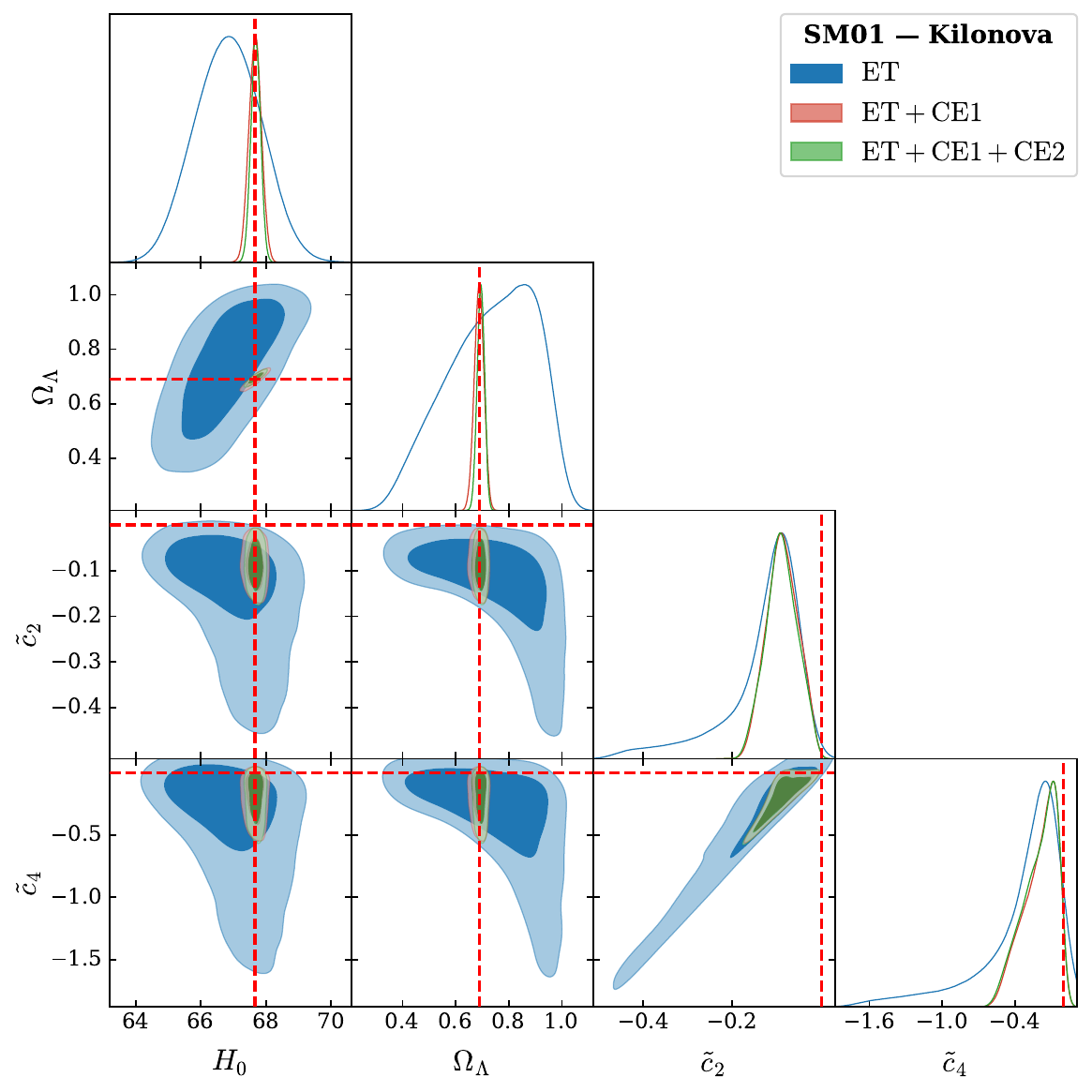}}\\[2pt]

\subfigure[\bf Prompt Emission]{\includegraphics[width=0.336\textwidth]{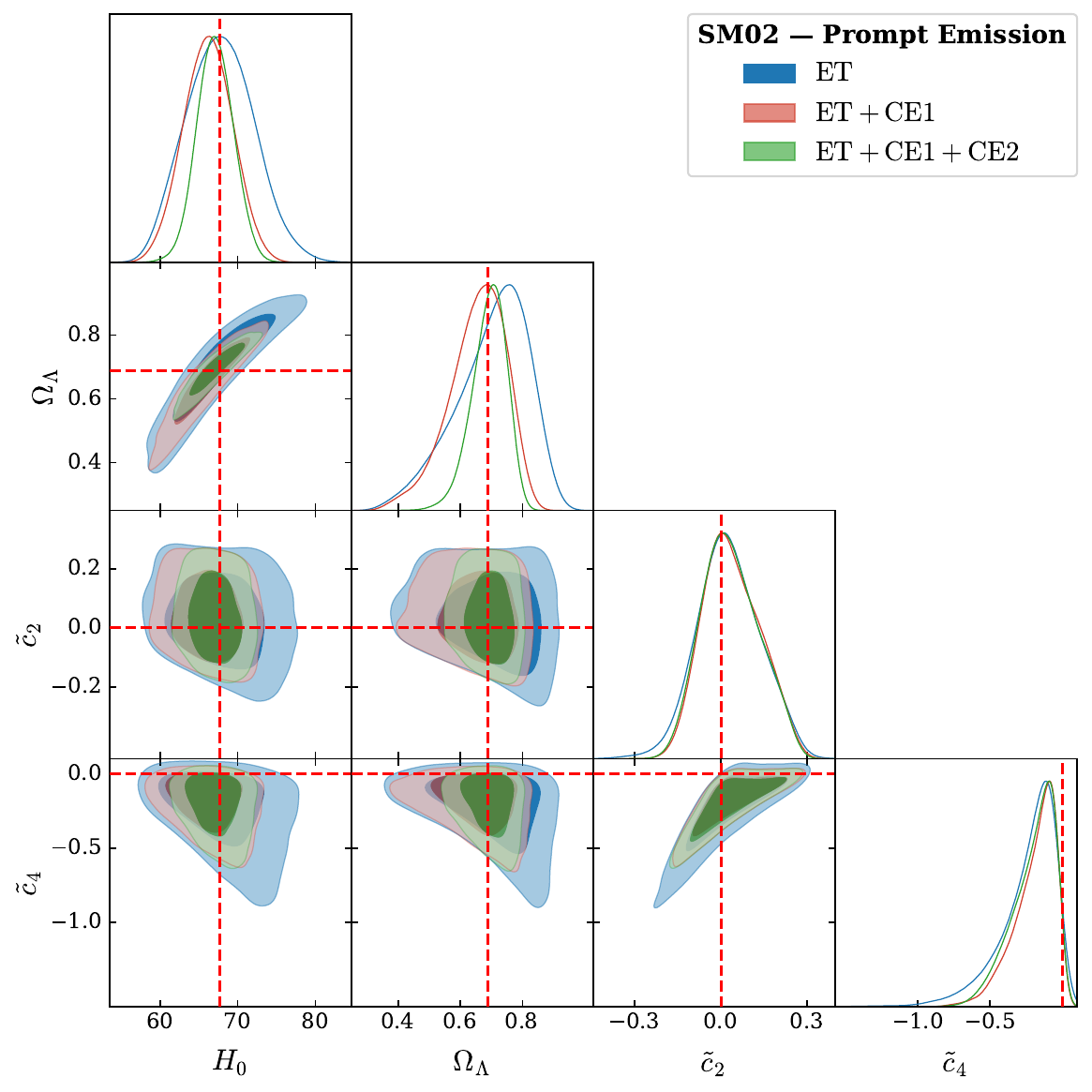}}\hspace{-5pt}
\subfigure[\bf Afterglow]{\includegraphics[width=0.336\textwidth]{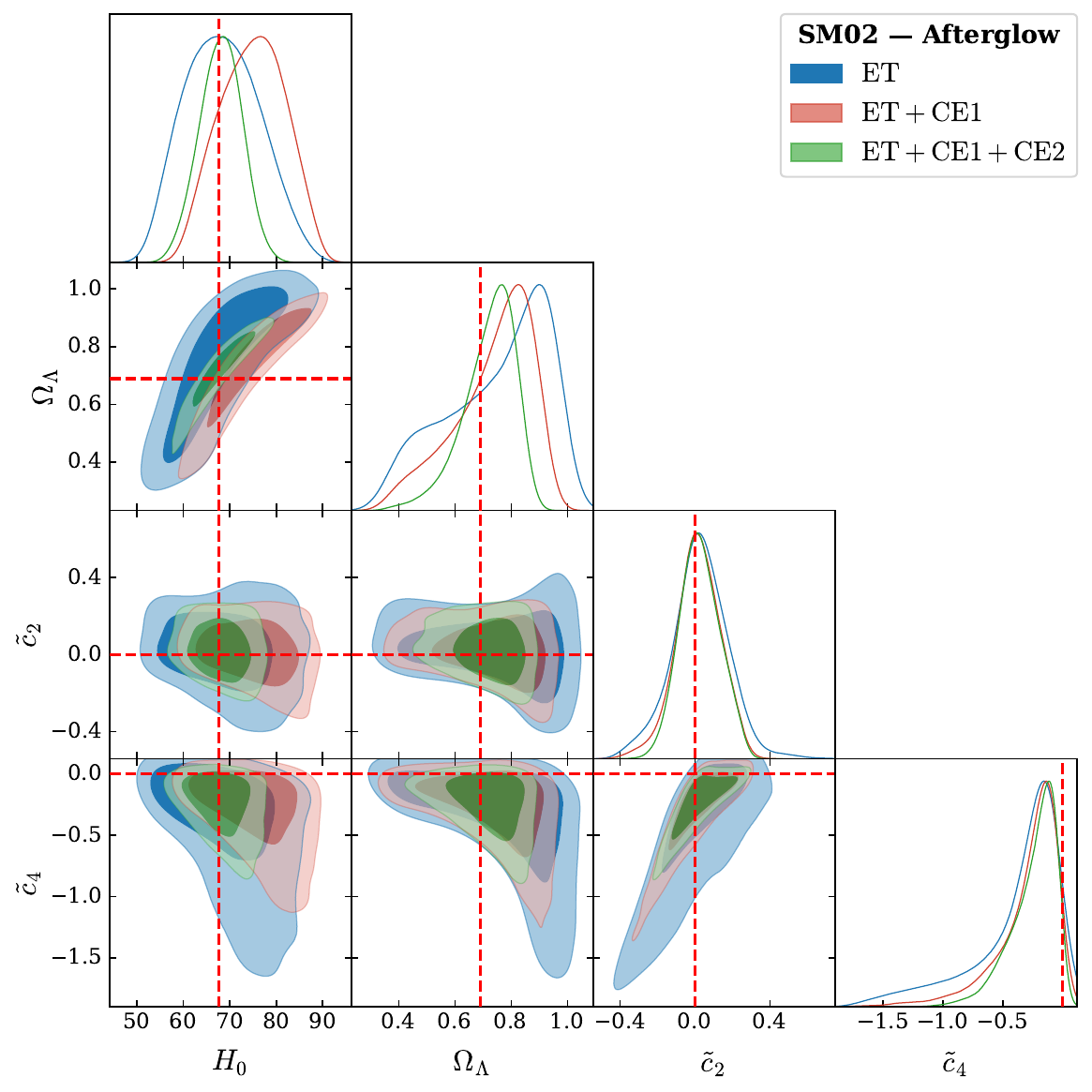}}\hspace{-5pt}
\subfigure[\bf Kilonova]{\includegraphics[width=0.336\textwidth]{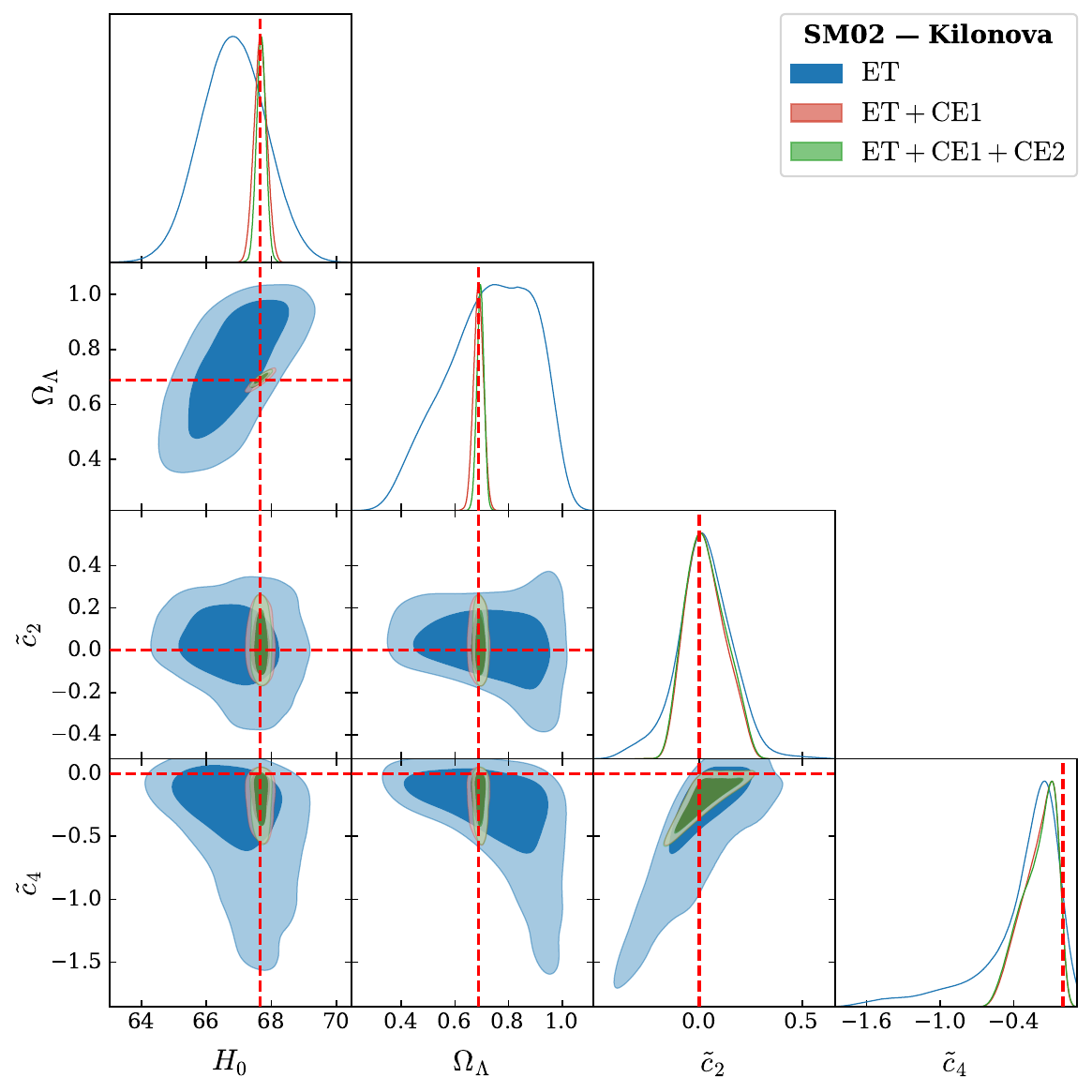}}

\caption{Posterior distributions (68\% and 95\% credible intervals) of the sampled parameters $(H_0,\,\Omega_\Lambda,\, \tilde{c}_2,\, \tilde{c}_4)$ for the Extended Cuscuton submodels SM01 and SM02 (rows) and the three EM counterpart channels: (a)~Prompt Emission, (b)~Afterglow, (c)~Kilonova (columns). Blue, red, and green contours correspond to the ET, ET+CE, and ET+2CE detector networks, respectively. Red dashed lines mark the fiducial $\Lambda$CDM values ($H_0=67.66\,\mathrm{km\,s^{-1}\,Mpc^{-1}}$, $\Omega_\Lambda=0.6889$, $\tilde{c}_2=\tilde{c}_4=0$).}
\label{fig:corner_SM01_02}
\end{figure*}

\begin{figure*}
\centering
{\includegraphics[width=0.336\textwidth]{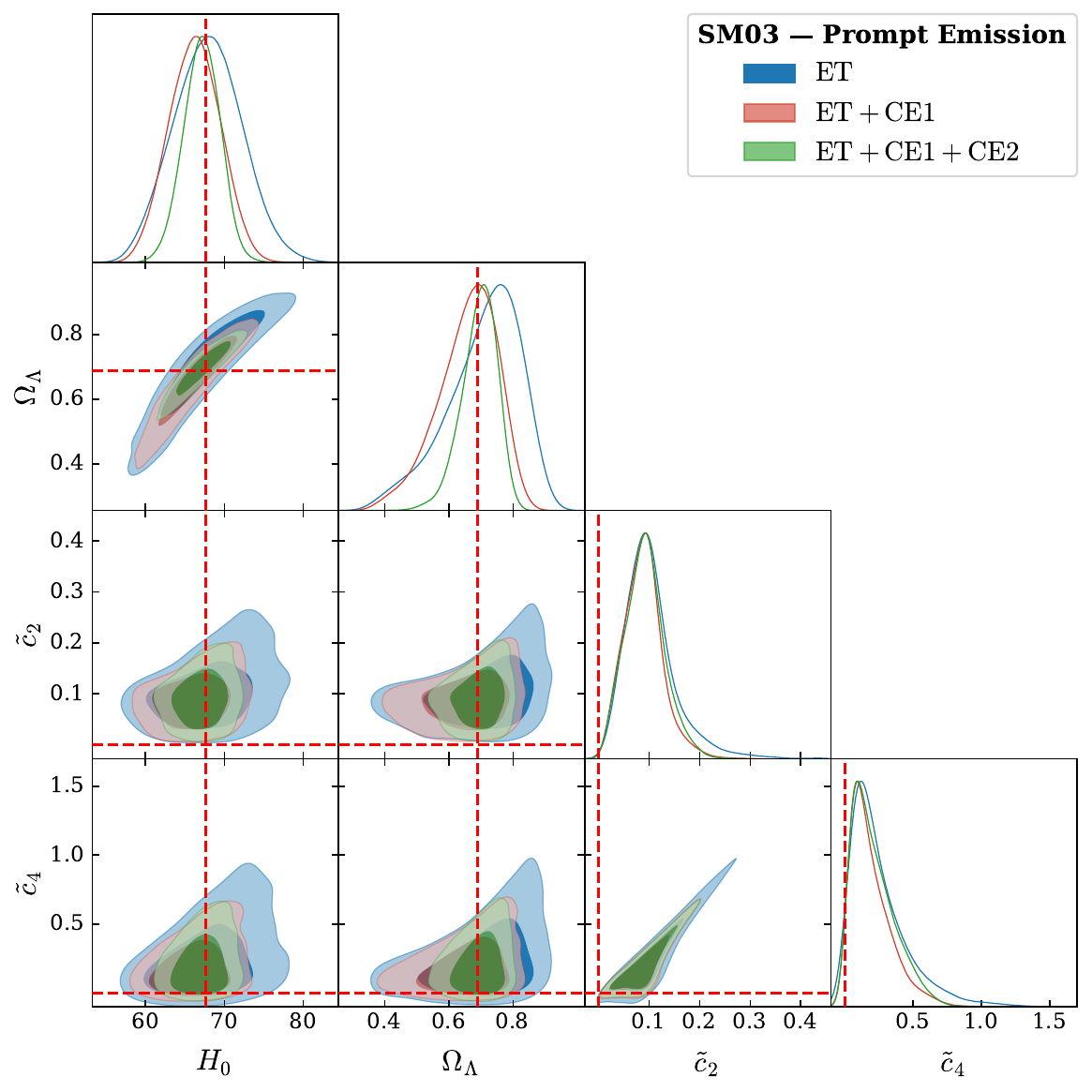}}\hspace{-5pt}
{\includegraphics[width=0.336\textwidth]{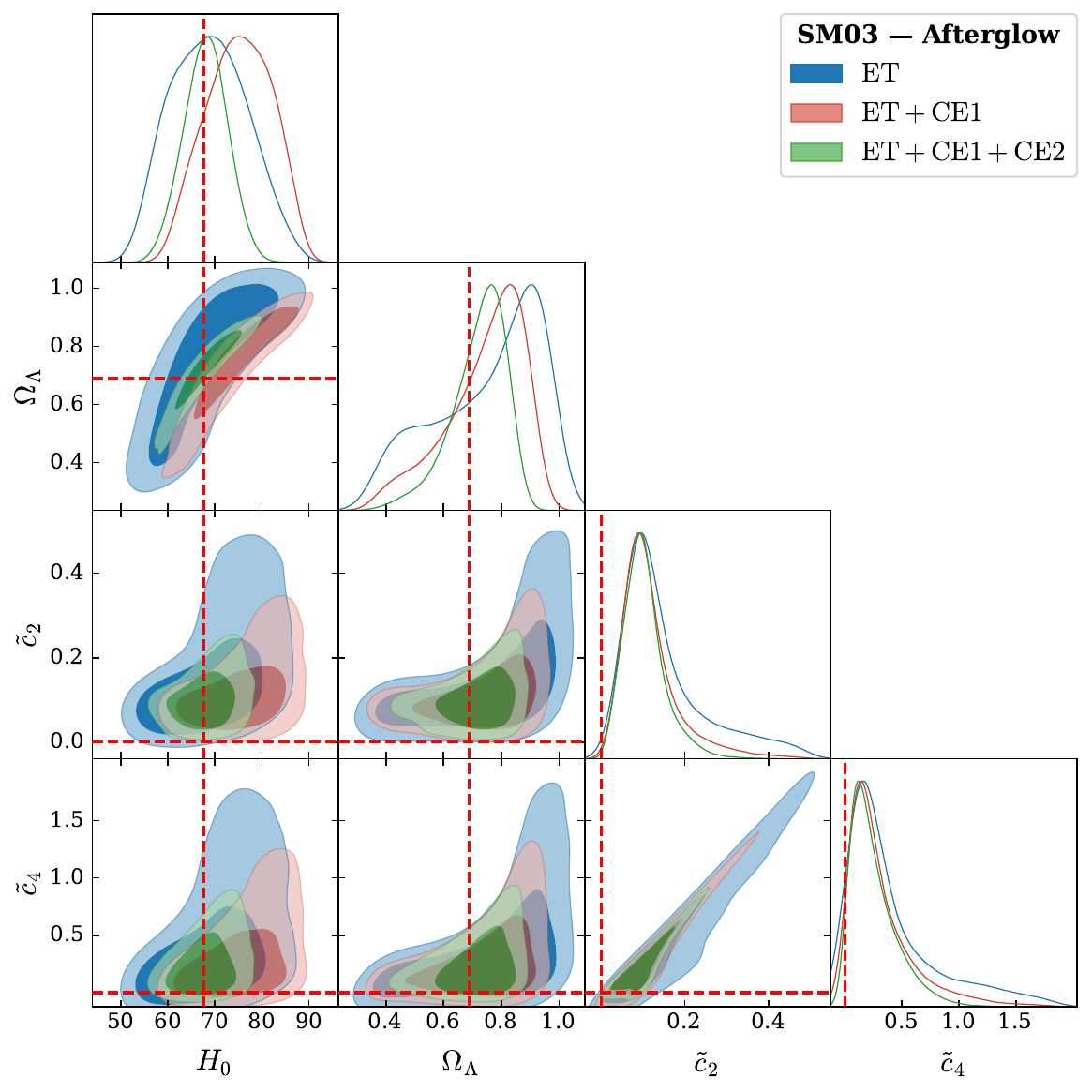}}\hspace{-5pt}
{\includegraphics[width=0.336\textwidth]{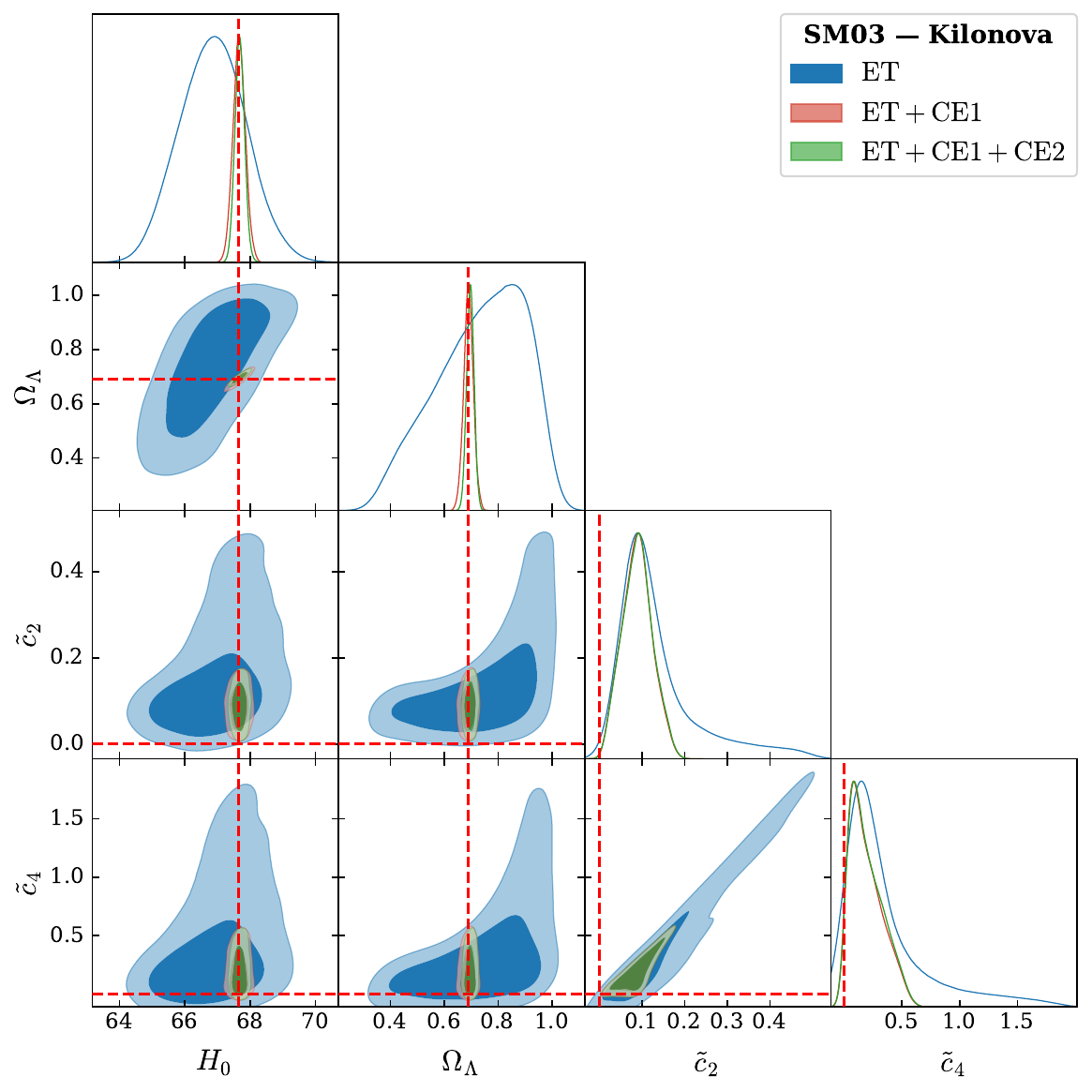}}\\[10pt]

\subfigure[\bf Prompt Emission]{\includegraphics[width=0.336\textwidth]{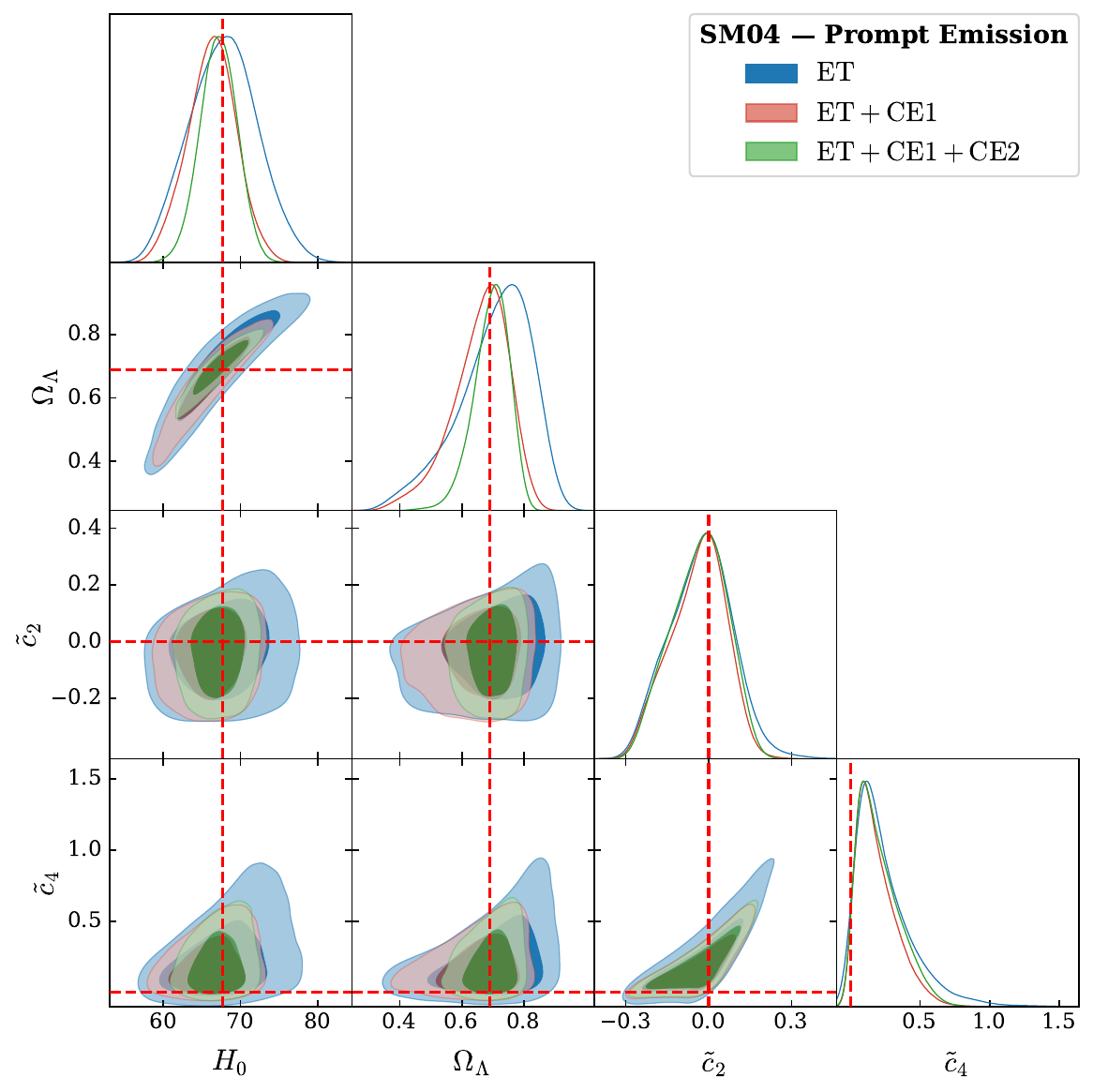}}\hspace{-5pt}
\subfigure[\bf Afterglow]{\includegraphics[width=0.336\textwidth]{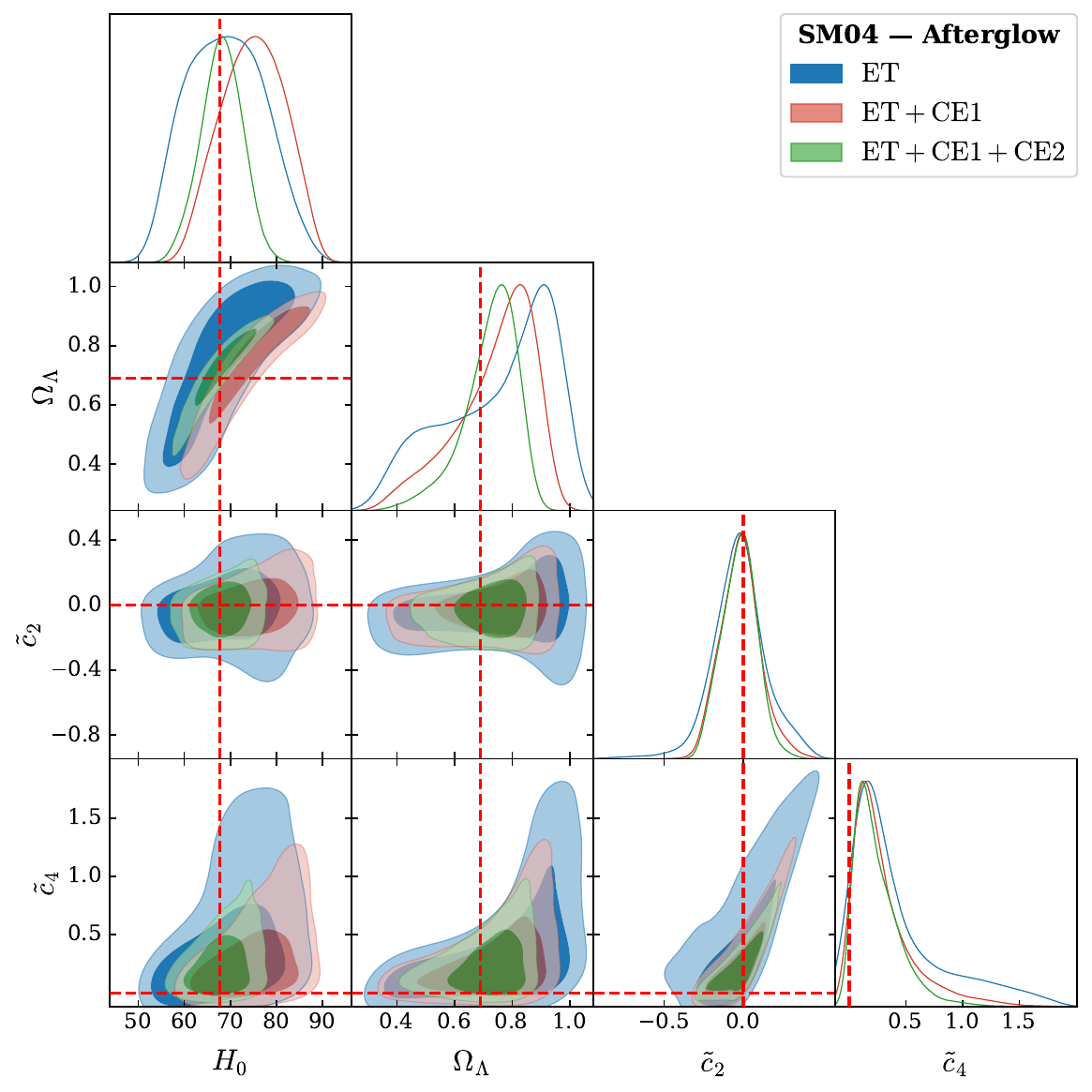}}\hspace{-5pt}
\subfigure[\bf Kilonova]{\includegraphics[width=0.336\textwidth]{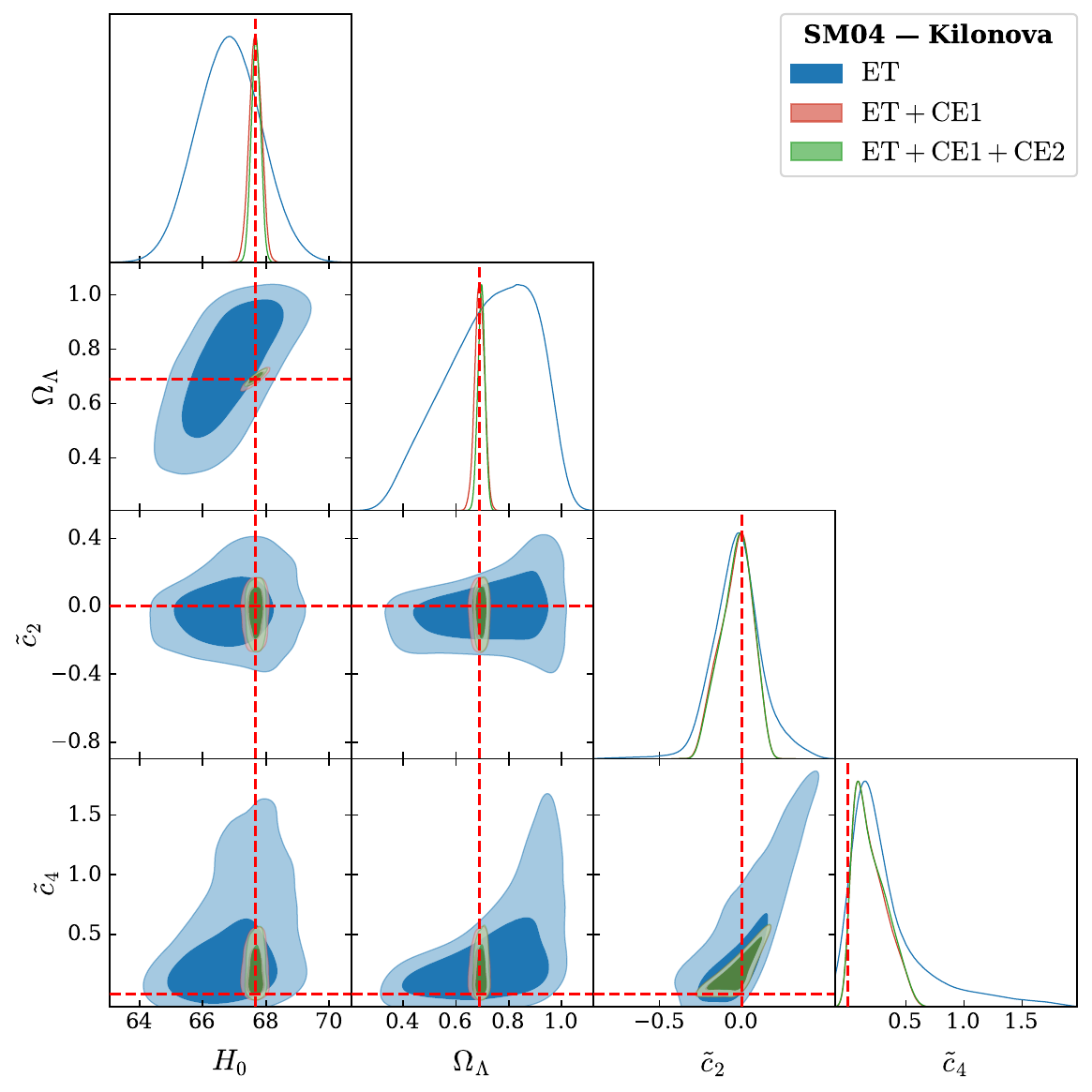}}

\caption{Posterior distributions (68\% and 95\% credible intervals) of the sampled parameters $(H_0,\,\Omega_\Lambda,\, \tilde{c}_2,\, \tilde{c}_4)$ for the Extended Cuscuton submodels SM03 and SM04 (rows) and the three EM counterpart channels: (a)~Prompt Emission, (b)~Afterglow, (c)~Kilonova (columns). Blue, red, and green contours correspond to the ET, ET+CE, and ET+2CE detector networks, respectively. Red dashed lines mark the fiducial $\Lambda$CDM values ($H_0=67.66\,\mathrm{km\,s^{-1}\,Mpc^{-1}}$, $\Omega_\Lambda=0.6889$, $\tilde{c}_2=\tilde{c}_4=0$).}
\label{fig:corner_SM03_04}
\end{figure*}

Figures~\ref{fig:corner_SM01_02} and~\ref{fig:corner_SM03_04} show the posterior distributions (68\% and 95\% credible intervals) of the sampled parameters $(H_0,\,\Omega_\Lambda,\, \tilde{c}_2,\, \tilde{c}_4)$ for the selected Extended Cuscuton submodels. In Fig.~\ref{fig:corner_SM01_02}, the first and second rows correspond to SM01 and SM02, respectively, while in Fig.~\ref{fig:corner_SM03_04} they correspond to SM03 and SM04. In both figures, the three columns correspond to the three electromagnetic counterpart channels: prompt emission, afterglow, and kilonova. The blue, red, and green contours correspond to the ET, ET+CE, and ET+2CE detector networks, respectively, while the red dashed lines mark the fiducial $\Lambda$CDM values, $H_0=67.66\,\mathrm{km\,s^{-1}\,Mpc^{-1}}$, $\Omega_\Lambda=0.6889$, and $\tilde{c}_2=\tilde{c}_4=0$.

First, no significant bias is encountered in the analysis. In all cases, the fiducial $\Lambda$CDM values are recovered within the inferred credible regions, confirming the consistency of the forecasting pipeline.

Second, despite the larger parameter space compared with flat $\Lambda$CDM, the relative error on the Hubble constant remains below $13.2\%$ in all cases and reaches a minimum of $0.21\%$ for SM02 with the ET+2CE kilonova catalogue. The corresponding relative error on $\Omega_\Lambda$ ranges from $1.87\%$ for SM02 with the ET+2CE kilonova catalogue to $27.2\%$ for SM01 with the ET afterglow catalogue.

Finally, the constraints on $\tilde{c}_2$, representing the linear contribution in the non-minimal coupling, and $\tilde{c}_4$, being the standard cuscuton mass term, are more model dependent. Keeping in mind that the fiducial $\Lambda$CDM limit corresponds to $\tilde{c}_2=\tilde{c}_4=0$, summarizing the constraints in terms of relative errors with respect to the posterior medians, the relative uncertainty on $\tilde{c}_2$ ranges from $39.3\%$ to $80.3\%$ for SM01, from $362\%$ to $697\%$ for SM02, from $41.0\%$ to $82.7\%$ for SM03, and from $364\%$ to $781\%$ for SM04. For $\tilde{c}_4$, the relative uncertainty ranges from $93.1\%$ to $149.8\%$ for SM01, from $88.8\%$ to $150.2\%$ for SM02, from $94.2\%$ to $154.4\%$ for SM03, and from $92.1\%$ to $162.6\%$ for SM04. 

These large fractional uncertainties should not be interpreted as a loss of absolute constraining power alone, but mainly reflect the fact that the posterior medians of $\tilde c_2$ in SM02 and SM04 lie close to the $\Lambda$CDM value.

\section{Discussions and conclusions}\label{conclusions}

We have studied a simple analytically tractable sector of the Extended Cuscuton model and tested it against both current low-redshift cosmological data and future mock standard-siren observations. The interest of the selected models lies in the possibility of describing non-dynamical dark energy through a minimal modification of the standard cosmological model within the post-GW170817/GRB170817A viable Horndeski sector. In this framework, there are only two tensor modes, which propagate at the speed of light, but the non-minimal coupling induces a running effective Planck mass and therefore a modified gravitational-wave luminosity distance. Standard sirens consequently probe the model through two complementary channels: the background expansion history and the modified amplitude damping of gravitational waves.

Currently, the combination of cosmic chronometers, Type-Ia supernovae, and BAO already restricts the models to a relatively narrow region of parameter space once the positivity of the effective gravitational coupling, Lunar Laser Ranging bounds, and Big Bang Nucleosynthesis constraints are imposed. 
Taking into account SM01, for the full CC$+$BAO$+$SN combination we obtain $\Omega_\Lambda = 0.73^{+0.01}_{-0.01}$ and $H_0 = 71.71^{+0.20}_{-0.31}$~km\,s$^{-1}$\,Mpc$^{-1}$. All four models agree on the background parameters to well within $1\sigma$; the model dependence is entirely absorbed by the shape parameters $\tc_2$ and $\tc_4$, which remain small ($|\tc_2| \lesssim 0.1$, $|\tc_4| \lesssim 0.3$) and stable across data combinations. The allowed solutions therefore lie close to the $\Lambda$CDM limit, as expected for a viable late-time modification of gravity subject to both cosmological and local constraints. The mild tension between the CC$+$SN and CC$+$BAO calibrations of $H_0$ ($72.50^{+0.81}_{-0.96}$ versus $69.20^{+0.86}_{-0.87}$ km\,s$^{-1}$\, Mpc$^{-1}$, for SM01) persists within this framework, indicating that the polynomial Extended Cuscuton model, while sufficiently flexible to remain observationally viable, does not by itself resolve the discrepancy between early- and late-time anchors of the distance scale. It is nonetheless already tightly constrained by current background probes; whether the small departures from $\Lambda$CDM encoded in $\tilde c_2$ and $\tilde c_4$ can be further probed is precisely the question addressed by our standard-siren forecasts.

The forecast analysis shows that 3G bright sirens can substantially sharpen this picture. As a first consistency check, fitting the fiducial flat $\Lambda$CDM model to the mock catalogues always recovers the input cosmology within the inferred credible regions, validating both the catalogue construction and the Bayesian inference methodology. Moving to the selected Extended Cuscuton submodels, the same qualitative trend is found throughout: the transition from ET to ET+CE and then to ET+2CE systematically improves the constraints, with the strongest performance generally obtained for the kilonova channel. This is expected, since the kilonova catalogues provide the largest number of well-localized events with secure redshift information, while the afterglow channel remains the least constraining because of its smaller effective sample size.

Even though the Extended Cuscuton models enlarge the parameter space with respect to flat $\Lambda$CDM, the cosmological parameters remain well constrained. Across the four submodels and the different mock datasets, the relative uncertainty on $H_0$ remains below $13.18\%$, reaching a minimum value of $0.21\%$ for SM02 with the KN catalogue in the ET+2CE network. The dark-energy density parameter is also efficiently recovered, with the relative uncertainty on $\Omega_\Lambda$ ranging from $1.87\%$ to $27.22\%$, again with the best constraint obtained for SM02 in the KN, ET+2CE configuration. These results show that the loss of precision associated with the enlarged parameter space is limited, especially for the most informative 3G network configurations.

The forecasts also preserve a clear hierarchy among the selected benchmark submodels. The pairs SM01/SM03 and SM02/SM04 behave very similarly, indicating that the sign of $\tilde c_1$ alone does not produce a large observational separation at the level of the background expansion and GW luminosity distance considered here. By contrast, the distinction between the two de Sitter branches is more relevant. The branches with $\varphi_{\rm dS}=0$ are generally more tightly constrained than those with $\varphi_{\rm dS}\neq0$, especially for the parameter $\tilde c_2$. Quantitatively, the relative uncertainty on $\tilde c_2$ ranges from $39.33\%$ to $80.28\%$ for SM01 and from $41.01\%$ to $82.73\%$ for SM03, while it increases to the range $362.07\%$--$697.06\%$ for SM02 and $363.79\%$--$781.25\%$ for SM04. This behaviour is a consequence of the fact that, in the non-vanishing asymptotic-scalar branch, the median value of $\tilde c_2$ remains close to zero, making the corresponding relative uncertainty large. Since $\tilde c_2$ is compatible with zero in SM02 and SM04, fractional uncertainties with respect to the posterior median are not a robust measure of constraining power. We therefore interpret these values mainly as an indication that the corresponding branch remains centred close to the $\Lambda$CDM limit. The parameter $\tilde c_4$, associated with the classical Cuscuton contribution, is less sharply determined in all submodels, with relative uncertainties of order unity: $93.14\%$--$149.78\%$ for SM01, $88.82\%$--$150.23\%$ for SM02, $94.21\%$--$154.42\%$ for SM03, and $92.07\%$--$162.55\%$ for SM04. This indicates that the mock standard-siren catalogues are mainly sensitive to the combinations of couplings that control the leading departures from the fiducial expansion history and from the standard tensor-amplitude damping.

From a physical perspective, 3G standard sirens can turn a minimally modified but tightly constrained dark-energy sector into a quantitatively testable target. The benchmark Extended Cuscuton models remain close to $\Lambda$CDM, as required by current data and local constraints, but they are not observationally inert. Future bright-siren catalogues can retain non-trivial sensitivity to the genuinely non-minimal couplings while simultaneously delivering precise measurements of $H_0$ and $\Omega_\Lambda$. This is precisely the regime in which standard sirens are most useful: not only as an independent way to reproduce a $\Lambda$CDM distance-redshift relation, but as a probe of whether a viable alternative theory remains distinguishable once both background evolution and tensor propagation effects are included.

Several extensions naturally follow from the present analysis. On the observational side, the most immediate step is a homogeneous update of the current-data analysis using the latest BAO compilations and the same pipeline adopted here. On the theoretical side, it will be important to move beyond the purely background-level treatment by including radiation consistently in early-time analyses and by studying the perturbation sector. In particular, a complete treatment should clarify the relation between the tensor coupling controlling $d_L^{\rm GW}$, the locally measured Newton constant constrained by Lunar Laser Ranging, and the effective scalar-sector coupling entering the growth of matter perturbations. It would also be worthwhile to revisit the broader analytically tractable subclass with non-zero $\tilde c_3$ and $\tilde c_5$, using methods specifically designed to diagnose and handle prior-dominated directions, such as profile-likelihood analyses or dedicated reparametrizations. Finally, extending the forecast to dark sirens and to more realistic multimessenger selection functions would provide a more complete assessment of the observational reach of future detector networks.

Overall, the results presented here support a clear conclusion: the Extended Cuscuton model provides a controlled and phenomenologically meaningful target for standard-siren cosmology beyond $\Lambda$CDM. Third-generation GW detector networks, especially when combined with electromagnetic counterpart observations, should be capable of testing this class of non-dynamical dark-energy models with significant precision.

\section*{Acknowledgements}

MM, DV, and SC are grateful for the support of Istituto Nazionale di Fisica Nucleare (INFN) iniziative specifiche MOONLIGHT-2, QGSKY, and TEONGRAV. MM thanks the University of Salamanca for the hospitality. IDM and R acknowledge support from the grant PID2024-158938NB-I00 funded by MCIN/AEI/10.13039/501100011033 and by ``ERDF A way of making Europe”, and from the grant SA097P24 funded by Junta de Castilla y León and by ``ERDF A way of making Europe”. We also acknowledge the use of the HPC facility Pegasus at IUCAA, Pune, India. This paper is based upon work from COST Action CA21136 -- Addressing observational tensions in cosmology with systematics and fundamental physics (CosmoVerse), supported by COST (European Cooperation in Science and Technology).

\bibliography{biblio}


\begin{table*}[b]
\caption{
Posterior median values and central $68.27\%$ credible intervals for SM01 and SM02 of the Extended Cuscuton model ($\tilde{c}_1=+1$, $\tilde{c}_3=\tilde{c}_5=0$). For each detector network, columns report the prompt-emission (PE), afterglow (AF), and kilonova (KN) channels. $\Omega_{m,0}$ and $\tilde{c}_6$ are derived parameters.
}
\centering
\renewcommand{\arraystretch}{2.0}
\setlength{\tabcolsep}{6pt}
\small
\begin{tabular}{c|c|ccc|ccc}
\hline
\multicolumn{2}{c|}{$\tilde{c}_1=+1,\enspace \tilde{c}_3=\tilde{c}_5=0$} & \multicolumn{3}{c|}{\textbf{SM01}} & \multicolumn{3}{c}{\textbf{SM02}} \\
\hline
\bf Parameter & \bf Network & \bf PE & \bf AF & \bf KN & \bf PE & \bf AF & \bf KN \\
\hline
\multirow{3}{*}{\boldmath$H_0$} & ET & $67.83^{+4.31}_{-4.50}$ & $68.31^{+9.03}_{-8.75}$ & $66.84^{+1.04}_{-1.04}$ & $67.74^{+4.39}_{-4.49}$ & $68.21^{+8.73}_{-8.34}$ & $66.83^{+1.00}_{-1.01}$ \\
 & ET+CE & $66.24^{+3.27}_{-3.24}$ & $75.14^{+7.05}_{-8.26}$ & $67.67^{+0.19}_{-0.20}$ & $66.20^{+3.23}_{-3.22}$ & $75.12^{+7.15}_{-7.94}$ & $67.66^{+0.20}_{-0.20}$ \\
 & ET+2CE & $67.11^{+2.42}_{-2.39}$ & $68.29^{+4.49}_{-4.73}$ & $67.68^{+0.15}_{-0.14}$ & $67.07^{+2.38}_{-2.30}$ & $68.18^{+4.57}_{-4.82}$ & $67.69^{+0.14}_{-0.14}$ \\
\hline
\multirow{3}{*}{\boldmath$\Omega_\Lambda$} & ET & $0.72^{+0.10}_{-0.14}$ & $0.80^{+0.15}_{-0.28}$ & $0.76^{+0.16}_{-0.20}$ & $0.72^{+0.10}_{-0.14}$ & $0.80^{+0.15}_{-0.27}$ & $0.75^{+0.17}_{-0.19}$ \\
 & ET+CE & $0.67^{+0.08}_{-0.10}$ & $0.77^{+0.11}_{-0.19}$ & $0.69^{+0.02}_{-0.02}$ & $0.66^{+0.08}_{-0.10}$ & $0.77^{+0.11}_{-0.19}$ & $0.69^{+0.02}_{-0.02}$ \\
 & ET+2CE & $0.70^{+0.05}_{-0.07}$ & $0.74^{+0.08}_{-0.11}$ & $0.70^{+0.01}_{-0.01}$ & $0.70^{+0.05}_{-0.06}$ & $0.74^{+0.07}_{-0.11}$ & $0.70^{+0.01}_{-0.01}$ \\
\hline
\multirow{3}{*}{\boldmath$\tilde{c}_2$} & ET & $-0.10^{+0.04}_{-0.05}$ & $-0.11^{+0.05}_{-0.13}$ & $-0.10^{+0.04}_{-0.10}$ & $0.02^{+0.12}_{-0.10}$ & $0.03^{+0.15}_{-0.14}$ & $0.02^{+0.14}_{-0.12}$ \\
 & ET+CE & $-0.09^{+0.04}_{-0.03}$ & $-0.10^{+0.04}_{-0.06}$ & $-0.09^{+0.04}_{-0.04}$ & $0.03^{+0.12}_{-0.09}$ & $0.02^{+0.12}_{-0.12}$ & $0.02^{+0.11}_{-0.09}$ \\
 & ET+2CE & $-0.09^{+0.04}_{-0.04}$ & $-0.10^{+0.04}_{-0.05}$ & $-0.09^{+0.04}_{-0.04}$ & $0.03^{+0.12}_{-0.09}$ & $0.02^{+0.12}_{-0.10}$ & $0.02^{+0.12}_{-0.09}$ \\
\hline
\multirow{3}{*}{\boldmath$\tilde{c}_4$} & ET & $-0.17^{+0.12}_{-0.24}$ & $-0.22^{+0.17}_{-0.50}$ & $-0.20^{+0.15}_{-0.40}$ & $-0.18^{+0.13}_{-0.24}$ & $-0.22^{+0.16}_{-0.49}$ & $-0.20^{+0.15}_{-0.38}$ \\
 & ET+CE & $-0.15^{+0.11}_{-0.17}$ & $-0.19^{+0.14}_{-0.30}$ & $-0.16^{+0.12}_{-0.18}$ & $-0.15^{+0.11}_{-0.18}$ & $-0.20^{+0.14}_{-0.32}$ & $-0.17^{+0.12}_{-0.18}$ \\
 & ET+2CE & $-0.17^{+0.12}_{-0.19}$ & $-0.19^{+0.14}_{-0.24}$ & $-0.16^{+0.12}_{-0.19}$ & $-0.17^{+0.12}_{-0.19}$ & $-0.19^{+0.14}_{-0.24}$ & $-0.17^{+0.12}_{-0.19}$ \\
\hhline{=|=|===|===}
\multirow{3}{*}{\boldmath$\tilde{c}_6$} & ET & $-0.02^{+0.04}_{-0.06}$ & $-0.02^{+0.04}_{-0.12}$ & $-0.02^{+0.04}_{-0.09}$ & $-0.02^{+0.04}_{-0.06}$ & $-0.02^{+0.05}_{-0.11}$ & $-0.02^{+0.04}_{-0.09}$ \\
 & ET+CE & $-0.01^{+0.03}_{-0.04}$ & $-0.02^{+0.04}_{-0.08}$ & $-0.01^{+0.04}_{-0.05}$ & $-0.01^{+0.03}_{-0.05}$ & $-0.02^{+0.04}_{-0.08}$ & $-0.02^{+0.04}_{-0.04}$ \\
 & ET+2CE & $-0.01^{+0.04}_{-0.05}$ & $-0.02^{+0.04}_{-0.06}$ & $-0.01^{+0.04}_{-0.05}$ & $-0.01^{+0.04}_{-0.05}$ & $-0.02^{+0.04}_{-0.06}$ & $-0.01^{+0.04}_{-0.05}$ \\
\hline
\multirow{3}{*}{\boldmath$\Omega_{m,0}$} & ET & $0.28^{+0.14}_{-0.10}$ & $0.20^{+0.28}_{-0.15}$ & $0.24^{+0.20}_{-0.16}$ & $0.28^{+0.14}_{-0.10}$ & $0.20^{+0.27}_{-0.15}$ & $0.25^{+0.19}_{-0.16}$ \\
 & ET+CE & $0.33^{+0.10}_{-0.08}$ & $0.23^{+0.19}_{-0.10}$ & $0.31^{+0.02}_{-0.02}$ & $0.33^{+0.10}_{-0.08}$ & $0.23^{+0.18}_{-0.10}$ & $0.31^{+0.02}_{-0.02}$ \\
 & ET+2CE & $0.30^{+0.07}_{-0.05}$ & $0.26^{+0.11}_{-0.08}$ & $0.30^{+0.01}_{-0.01}$ & $0.30^{+0.06}_{-0.05}$ & $0.26^{+0.11}_{-0.07}$ & $0.30^{+0.01}_{-0.01}$ \\
\hline
\end{tabular}

\label{tab:ec_sm01_sm02_5yr}
\end{table*}


\begin{table*}[ht!]
\caption{
Posterior median values and central $68.27\%$ credible intervals for SM03 and SM04 of the Extended Cuscuton model ($\tilde{c}_1=-1$, $\tilde{c}_3=\tilde{c}_5=0$). For each detector network, columns report the prompt-emission (PE), afterglow (AF), and kilonova (KN) channels. $\Omega_{m,0}$ and $\tilde{c}_6$ are derived parameters.
}
\centering
\renewcommand{\arraystretch}{2.0}
\setlength{\tabcolsep}{6pt}
\small
\begin{tabular}{c|c|ccc|ccc}
\hline
\multicolumn{2}{c|}{$\tilde{c}_1=-1,\enspace \tilde{c}_3=\tilde{c}_5=0$} & \multicolumn{3}{c|}{\textbf{SM03}} & \multicolumn{3}{c}{\textbf{SM04}} \\
\hline
\bf Parameter & \bf Network & \bf PE & \bf AF & \bf KN & \bf PE & \bf AF & \bf KN \\
\hline
\multirow{3}{*}{\boldmath$H_0$} & ET & $67.91^{+4.30}_{-4.52}$ & $68.68^{+8.78}_{-8.79}$ & $66.87^{+1.01}_{-1.04}$ & $67.86^{+4.32}_{-4.66}$ & $68.85^{+9.01}_{-9.13}$ & $66.84^{+1.02}_{-1.01}$ \\
 & ET+CE & $66.28^{+3.33}_{-3.38}$ & $75.20^{+7.57}_{-7.77}$ & $67.66^{+0.19}_{-0.18}$ & $66.40^{+2.98}_{-3.30}$ & $75.03^{+7.23}_{-7.64}$ & $67.67^{+0.20}_{-0.19}$ \\
 & ET+2CE & $67.15^{+2.33}_{-2.37}$ & $68.05^{+4.67}_{-4.85}$ & $67.68^{+0.15}_{-0.14}$ & $67.15^{+2.40}_{-2.40}$ & $68.18^{+4.53}_{-4.67}$ & $67.68^{+0.15}_{-0.15}$ \\
\hline
\multirow{3}{*}{\boldmath$\Omega_\Lambda$} & ET & $0.73^{+0.10}_{-0.14}$ & $0.81^{+0.14}_{-0.29}$ & $0.76^{+0.16}_{-0.20}$ & $0.72^{+0.10}_{-0.14}$ & $0.81^{+0.14}_{-0.29}$ & $0.75^{+0.16}_{-0.20}$ \\
 & ET+CE & $0.67^{+0.08}_{-0.11}$ & $0.78^{+0.10}_{-0.18}$ & $0.69^{+0.02}_{-0.02}$ & $0.67^{+0.08}_{-0.11}$ & $0.78^{+0.10}_{-0.18}$ & $0.69^{+0.02}_{-0.02}$ \\
 & ET+2CE & $0.70^{+0.05}_{-0.06}$ & $0.74^{+0.08}_{-0.11}$ & $0.70^{+0.01}_{-0.01}$ & $0.70^{+0.05}_{-0.06}$ & $0.74^{+0.08}_{-0.11}$ & $0.70^{+0.01}_{-0.01}$ \\
\hline
\multirow{3}{*}{\boldmath$\tilde{c}_2$} & ET & $0.10^{+0.05}_{-0.04}$ & $0.11^{+0.13}_{-0.05}$ & $0.10^{+0.10}_{-0.04}$ & $-0.02^{+0.11}_{-0.13}$ & $-0.02^{+0.15}_{-0.16}$ & $-0.03^{+0.13}_{-0.14}$ \\
 & ET+CE & $0.09^{+0.03}_{-0.04}$ & $0.10^{+0.07}_{-0.04}$ & $0.09^{+0.04}_{-0.04}$ & $-0.03^{+0.09}_{-0.12}$ & $-0.02^{+0.12}_{-0.13}$ & $-0.03^{+0.09}_{-0.12}$ \\
 & ET+2CE & $0.09^{+0.04}_{-0.04}$ & $0.10^{+0.05}_{-0.04}$ & $0.09^{+0.04}_{-0.04}$ & $-0.03^{+0.10}_{-0.12}$ & $-0.02^{+0.11}_{-0.12}$ & $-0.03^{+0.09}_{-0.11}$ \\
\hline
\multirow{3}{*}{\boldmath$\tilde{c}_4$} & ET & $0.18^{+0.25}_{-0.13}$ & $0.23^{+0.53}_{-0.17}$ & $0.20^{+0.42}_{-0.15}$ & $0.18^{+0.25}_{-0.13}$ & $0.23^{+0.58}_{-0.17}$ & $0.20^{+0.41}_{-0.15}$ \\
 & ET+CE & $0.15^{+0.19}_{-0.11}$ & $0.20^{+0.34}_{-0.14}$ & $0.16^{+0.19}_{-0.12}$ & $0.15^{+0.18}_{-0.11}$ & $0.20^{+0.31}_{-0.15}$ & $0.16^{+0.18}_{-0.12}$ \\
 & ET+2CE & $0.17^{+0.20}_{-0.12}$ & $0.19^{+0.26}_{-0.14}$ & $0.16^{+0.19}_{-0.12}$ & $0.17^{+0.20}_{-0.12}$ & $0.19^{+0.24}_{-0.13}$ & $0.17^{+0.19}_{-0.12}$ \\
\hhline{=|=|===|===}
\multirow{3}{*}{\boldmath$\tilde{c}_6$} & ET & $0.02^{+0.06}_{-0.04}$ & $0.02^{+0.13}_{-0.05}$ & $0.02^{+0.10}_{-0.04}$ & $0.02^{+0.06}_{-0.04}$ & $0.02^{+0.14}_{-0.05}$ & $0.02^{+0.10}_{-0.04}$ \\
 & ET+CE & $0.01^{+0.05}_{-0.03}$ & $0.02^{+0.08}_{-0.04}$ & $0.01^{+0.05}_{-0.04}$ & $0.01^{+0.05}_{-0.04}$ & $0.02^{+0.08}_{-0.04}$ & $0.01^{+0.05}_{-0.04}$ \\
 & ET+2CE & $0.02^{+0.05}_{-0.04}$ & $0.02^{+0.07}_{-0.04}$ & $0.02^{+0.05}_{-0.04}$ & $0.02^{+0.05}_{-0.04}$ & $0.02^{+0.06}_{-0.04}$ & $0.02^{+0.05}_{-0.04}$ \\
\hline
\multirow{3}{*}{\boldmath$\Omega_{m,0}$} & ET & $0.28^{+0.14}_{-0.10}$ & $0.19^{+0.29}_{-0.15}$ & $0.24^{+0.20}_{-0.16}$ & $0.28^{+0.14}_{-0.10}$ & $0.19^{+0.30}_{-0.14}$ & $0.25^{+0.20}_{-0.17}$ \\
 & ET+CE & $0.33^{+0.11}_{-0.08}$ & $0.22^{+0.18}_{-0.10}$ & $0.31^{+0.02}_{-0.02}$ & $0.33^{+0.11}_{-0.08}$ & $0.23^{+0.18}_{-0.10}$ & $0.31^{+0.02}_{-0.02}$ \\
 & ET+2CE & $0.30^{+0.06}_{-0.05}$ & $0.26^{+0.11}_{-0.08}$ & $0.30^{+0.01}_{-0.01}$ & $0.30^{+0.06}_{-0.05}$ & $0.26^{+0.11}_{-0.08}$ & $0.30^{+0.01}_{-0.01}$ \\
\hline
\end{tabular}
\label{tab:ec_sm03_sm04_5yr}
\end{table*}

\end{document}